\documentclass{ar-1col-S2O}

\usepackage[comma]{natbib}
\usepackage{url}

\usepackage{amsmath}
\usepackage{amssymb}     
\usepackage{pifont}      
\usepackage{booktabs}    
\usepackage{graphicx}
\usepackage{subcaption}

\jname{Annual Review of Astronomy and Astrophysics}
\jvol{AA}
\jyear{2025}
\doi{10.1146/}

\begin{document}

\markboth{Y.-S. Ting}{Deep Learning in Astrophysics}

\title{Deep Learning in Astrophysics}

\author{Yuan-Sen Ting,$^{1,2,3}$
\affil{$^1$Department of Astronomy, The Ohio State University, Columbus, Ohio, USA; email: ting.74@osu.edu}
\affil{$^2$Center for Cosmology and AstroParticle Physics (CCAPP), The Ohio State University, Columbus, Ohio, USA}
\affil{$^3$Max-Planck-Institut f\"ur Astronomie, Heidelberg, Germany}
}

\begin{abstract}
Deep learning has generated diverse perspectives in astronomy, with ongoing discussions between proponents and skeptics motivating this review. We examine how neural networks complement classical statistics, extending our data analytical toolkit for modern surveys. Astronomy offers unique opportunities through encoding physical symmetries, conservation laws, and differential equations directly into architectures, creating models that generalize beyond training data. Yet challenges persist as unlabeled observations number in billions while confirmed examples with known properties remain scarce and expensive. This review demonstrates how deep learning incorporates domain knowledge through architectural design, with built-in assumptions guiding models toward physically meaningful solutions. We evaluate where these methods offer genuine advances versus claims requiring careful scrutiny. 
$\bullet$~Neural architectures overcome bias-variance trade-offs among scalability, expressivity, and data efficiency by encoding physical symmetries and conservation laws into network structure, enabling learning from limited labeled data.
$\bullet$~Simulation-based inference and anomaly detection extract information from complex, non-Gaussian distributions where analytical likelihoods fail, enabling field-level cosmological analysis and systematic discovery of rare phenomena. 
$\bullet$~Multiscale neural modeling bridges resolution gaps in astronomical simulations, learning effective subgrid physics from expensive high-fidelity runs to enhance large-volume calculations where direct computation remains prohibitive.
$\bullet$~Emerging paradigms—reinforcement learning for telescope operations, foundation models learning from minimal examples, and large language model agents for research automation—show promise though are still developing in astronomical applications.
\end{abstract}


\maketitle

\tableofcontents

\vspace{-1cm}

\section{INTRODUCTION}

Astronomy has entered an unprecedented data-rich era. In response, astronomers have increasingly turned to machine learning, particularly deep learning techniques that promise to extract patterns from vast, high-dimensional datasets. The rapid adoption of these techniques reflects both necessity and opportunity—we need scalable methods to process modern surveys, and neural networks offer capabilities that classical approaches cannot match.

Yet this transformation has elicited varied responses across our community. Some researchers embrace deep learning as transformative, whereas others remain skeptical. These divergent perspectives can obscure a nuanced reality: We are witnessing genuine methodological advances that require careful evaluation. This discussion echoes historical patterns in artificial intelligence (AI). AI has experienced cycles of elevated expectations and subsequent recalibration---AI winters and AI springs. During periods of enthusiasm, terminology proliferates; when results fall short of initial promises, terminology evolves to distance new work from past failures. This evolution has created some ambiguity as terms like AI, machine learning, and deep learning are used interchangeably. Such imprecision can both obscure meaningful distinctions and allow enthusiasm to outpace careful validation.

The truth lies between extremes. Neural network-based approaches offer genuine advances in data modeling---which we explore throughout this review. Some developments represent capabilities beyond what we previously achieved. Yet we must learn from historical patterns of elevated expectations in this rapidly evolving field. This review aims to provide astronomers with a clear-eyed assessment: what deep learning genuinely offers our field, where and why it excels, and how we can leverage these tools for scientific discovery.

\subsection{What Is Machine Learning}
\label{subsection:what-is-ml}

\begin{figure}[h]
\centering
\includegraphics[width=\textwidth]{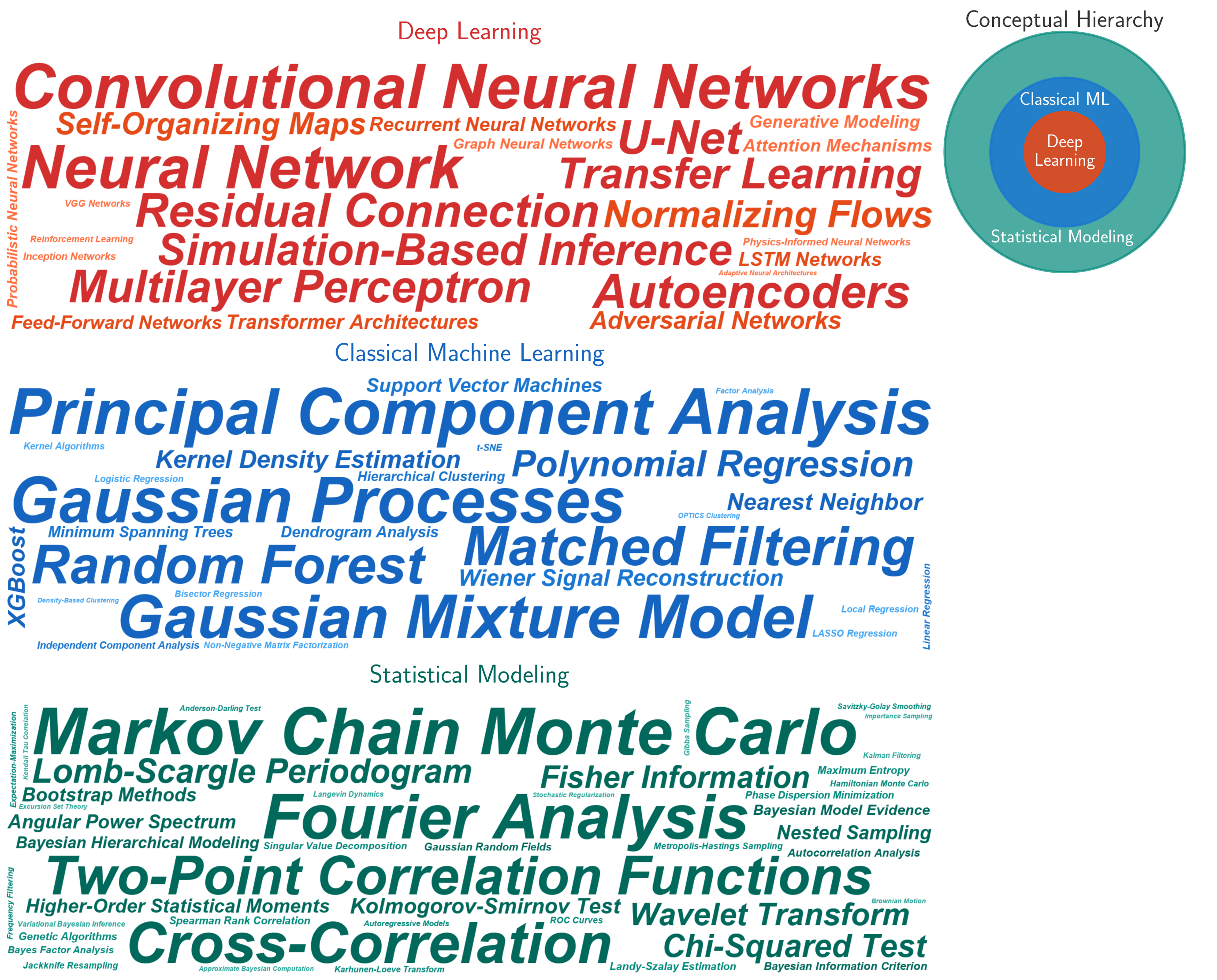}
\caption{Taxonomic hierarchy of statistical methods in astronomy. We adopt a nested framework in which machine learning encompasses methods that extract patterns from data to make predictions, with deep learning (neural network-based approaches) forming a specialized subset. (\textit{a}) Deep learning methods (red). (\textit{b}) Classical machine learning techniques (blue). (\textit{c}) Statistical methods underlying both approaches (green). Word sizes reflect prevalence in the astronomy literature as of July 2025. Abbreviations: LASSO, least absolute shrinkage and selection operator; LSTM, long short-term memory; ML, machine learning; OPTICS, ordering points to identify the clustering structure; ROC, receiver operating characteristic; t-SNE, t-distributed stochastic neighbor embedding; VGG, visual geometry group; XGBoost, extreme gradient boosting.}
\label{fig1}
\end{figure}

We define machine learning as algorithms constructing models from data that generalize to new observations. This involves learning patterns from data to make predictions (i.e., generalizations) or discover structure in unseen data. Through this lens, models can be built with or without neural networks. We define deep learning as neural network-based machine learning models. Machine learning models that are not neural network-based but commonly used in astronomy (see \textbf{Figure~\ref{fig1}}) we call classical machine learning \citep{Bishop2006,Ting2025a}. Statistical techniques that are relevant but not machine learning algorithms themselves form an outer circle in this nested hierarchy.

Machine learning shares the same goal as traditional physical modeling: understanding nature through quantitative analysis. The philosophical differences between ``theory-driven'' and ``data-driven'' approaches matter less than their shared aim of extracting meaningful patterns from nature. Statistical modeling—of which machine learning is a part—has been integral to astronomical discovery. For instance, linear regression, one of the simplest machine learning algorithms, has identified fundamental relationships like the M-$\sigma$ relation between supermassive black hole masses and bulge velocity dispersions \citep{Ferrarese2000,Gebhardt2000} and the Kennicutt-Schmidt relation linking gas surface density to star-formation rate \citep{Kennicutt1998}.

But as data volumes explode, the choice of method becomes critical. Astronomical applications demand more than what industrial AI offers; not only prediction accuracy but also physical understanding, uncertainty quantification, and discovery potential are needed. This unique need shapes three critical properties any astronomical machine learning approach must balance:

\begin{marginnote}[]
\entry{Generalizations}{extrapolating to new physical regimes and phenomena beyond training distribution, distinguishing true generalization from mere interpolation}
\end{marginnote}

\begin{itemize}
\item \textbf{Scalability}: Training models efficiently on large datasets and deploying them for inference at scale. Surveys now routinely produce billions of sources where decades ago they produced thousands.

\item \textbf{Generalizability}: The ability to extrapolate beyond training data---essential for scientific inquiry. Models must handle unexplored regimes and new phenomena while bridging synthetic models with observed data. True generalizability requires robustness to domain shifts---systematic differences between training and application contexts, such as models trained on simulations encountering real observations---while maintaining sensitivity to anomalies that could represent new physics.

\begin{marginnote}[]
\entry{Domain shift}{systematic differences between training and deployment data distributions; common astronomy examples include simulations to observations, different instruments, and varying observation selections}
\end{marginnote}

\item \textbf{Data efficiency}: The ability to learn from limited labeled examples. Spectroscopic confirmation requires telescope time that photometry does not; space-based follow-up costs orders of magnitude more per target; expert classifications demand specialized knowledge that cannot scale. This creates an asymmetry—vast unlabeled datasets but scarce confirmed labels in astronomy.
\end{itemize}

\subsection{Classical Machine Learning}
\label{subsection:classical-ml}

Classical machine learning methods each excel in different aspects of the three key properties (scalability, data efficiency, and expressivity), but none simultaneously achieve all three. Linear regression scales well and learns from minimal data \citep{Bishop2006,Ting2025a} but lacks expressivity for complex patterns. Gaussian processes offer principled uncertainty quantification but struggle with computational tractability at scale \citep{Rasmussen2006}. Random forests balance efficiency and expressivity \citep{Breiman2001} but cannot extrapolate beyond training data ranges. These complementary strengths and weaknesses motivate the development of physics-informed deep learning approaches that aim to unify all three properties (see \textbf{Supplemental Text, Section A}).

The root of these trade-offs lies in the curse of dimensionality. Computational requirements grow as $O(n^d)$ where $n$ is resolution per dimension and $d$ is dimensionality \citep[see][their chapter 5]{Bellman1961}. In high-dimensional spaces, data points become isolated islands. Random forests, operating as adaptive nearest-neighbor algorithms \citep{Lin2006}, cannot extrapolate because neighboring points become vanishingly rare. High-order polynomials might achieve zero training error yet fail catastrophically on new observations---interpolating perfectly between training points while oscillating wildly in unexplored regions. This paradox reveals that increased expressivity does not guarantee better performance. Classical methods thus face an inherent limitation: They cannot simultaneously achieve scalability, expressivity, and data efficiency in high-dimensional spaces.

\begin{marginnote}[]
\entry{Curse of dimensionality}{exponential growth in data requirements as dimensions increase; data points become isolated in high-dimensional spaces, requiring exponentially more samples}
\end{marginnote}

\subsection{Why Modern Surveys Demand New Methods}
\label{subsection:why-modern-surveys}

The inherent limitations of classical methods matter because modern surveys reveal phenomena invisible to simpler analyses. Classical approaches focus on dominant statistical features like two-point correlation functions capturing Gaussian fluctuations \citep{Peebles1980}. With billions of sources, however, surveys enable detection of information in higher-order statistics and rare events. In cosmology, this motivates exploration beyond traditional two-point statistics to probe primordial non-Gaussianity and nonlinear structure formation \citep{Bernardeau2002}. Time-domain astronomy exemplifies this progression: asteroseismology evolved from fitting individual oscillation frequencies to globally characterizing entire power spectra, revealing deep stellar structure through acoustic glitches and mixed modes \citep{Garcia2019}.

Expressivity and generalizability also matter for outlier detection, where many breakthroughs emerge. Fast radio bursts---millisecond pulses fitting no known category---emerged from big data analysis \citep{Lorimer2007}. Recent discoveries continue this pattern: Gaia detected stellar-mass black holes in wide orbits \citep{El-Badry2023}, Pan-STARRS1 (Panoramic Survey Telescope \& Rapid Response System) discovered 'Oumuamua's interstellar trajectory \citep{Meech2017}, time-domain surveys identified fast blue optical transients like AT 2018cow \citep{Prentice2018}, and surveys revealed extreme trans-Neptunian objects like 2017 OF201 \citep{Cheng2025}. The scale intensifies these challenges. For example, the Vera C. Rubin Observatory will detect 40 billion sources with 800 visits per location \citep{Ivezic2019}, and Euclid will image 1.5 billion galaxies \citep{Euclid2025}; spectroscopic surveys like Large Sky Area Multi-Object Fibre Spectroscopic Telescope (LAMOST) have collected 11.5 million spectra \citep{Abdurrouf2022}, and the Dark Energy Spectroscopic Instrument (DESI) Legacy Imaging Surveys will obtain 36 million galaxy spectra \citep{DESI2024}. We need models flexible enough to capture unexpected patterns and scalable enough to search datasets systematically.

\subsection{Deep Learning as a New Direction}

\label{subsection:deep-learning-direction}
The limitations of classical methods---their inability to simultaneously achieve scalability, expressivity, and data efficiency---create a bottleneck for modern astronomy. Neural networks offer an alternative approach. They provide scalability through graphics processing unit computation and expressivity for learning complex relationships, even in high dimensions. The universal approximation theorem proved feedforward networks with sufficient neurons can approximate any continuous function to arbitrary accuracy \citep{Cybenko1989,Hornik1989}.

In the early days of deep learning, this theoretical power came at a practical cost. Although theoretically capable of fitting any continuous function, achieving this fitting required massive training datasets. Critics articulated what appeared to be a fundamental barrier: the bias-variance dilemma \citep{Geman1992}. High bias means underfitting---oversimplified models missing real patterns. High variance means overfitting---like fitting passes through every point but the passes oscillate wildly between them. Neural networks seemed trapped---their expressivity reduced bias but amplified variance, making them sensitive to training examples.

\begin{marginnote}[]
\entry{Universal approximation theorem}{mathematical proof that feedforward neural networks with sufficient neurons can approximate any continuous function to arbitrary accuracy}
\end{marginnote}

Recent developments have challenged this understanding. The bias-variance dilemma proved less constraining than predicted. A key theoretical insight helps explain why: Neural networks with infinite width converge to Gaussian processes \citep{Neal1996,Williams1996,Lee2017}. Just as the central limit theorem ensures sums of random variables converge to Gaussians, infinitely wide networks---through averaging over many neurons---become equivalent to Gaussian processes with specific kernels. This connection reveals why overparameterized networks generalize better than simple polynomials of comparable flexibility \citep{Zhang2016,Belkin2019}.

Yet expressivity alone does not solve astronomy's challenges---flexible models still require substantial data. Most importantly for astronomy, the machine learning community discovered architectural innovations that improve data efficiency while enhancing interpretability. As we explore in Section \ref{section:methodological-foundations}, these advances incorporate inductive biases---domain knowledge encoded directly into network structure. Although neural networks have been applied in astronomy since the late 1980s \citep[e.g.,][]{Adorf1988,Angel1990}, their sophistication has transformed dramatically (see \textbf{Supplemental Text, Section B} for detailed historical evolution). Modern specialized architectures now encode physical assumptions directly. Section \ref{section:methodological-foundations} examines the theoretical foundations enabling these architectures to encode astronomical knowledge, whereas Section \ref{section:cross-cutting-techniques} demonstrates their application to practical challenges.

\section{METHODOLOGICAL FOUNDATIONS}
\label{section:methodological-foundations}

Deep learning achieves its success through architectural innovations, but understanding why these work requires examining a fundamental challenge in scientific inquiry: How can we justify inference from the observed to the unobserved? This challenge has philosophical roots in David Hume's problem of induction \citep{Hume1748}. Tom Mitchell formalized the answer, defining bias as ``any basis for choosing one generalization over another, other than strict consistency with the observed training instances'' \citep{Mitchell1980}. Mitchell proved that without such biases, learning algorithms cannot outperform memorization.

\begin{marginnote}[]
\entry{Inductive biases}{assumptions guiding learning algorithms toward solutions; encoded through architectural choices, physics constraints, or symmetries rather than explicit functional forms}
\end{marginnote}

In machine learning, we call these biases \textit{inductive biases}---the built-in preferences that guide algorithms toward certain solutions over others. These are the assumptions we embed into our models that enable them to generalize beyond the training data and learn underlying patterns. In astronomy, we constantly employ such biases. When fitting a power law to the M-$\sigma$ relation, we assume the underlying physics produces this specific functional form rather than exploring all possible curves. When modeling galaxy profiles with Sérsic functions or spectral lines with Voigt profiles, we constrain our search to particular mathematical families based on physical reasoning. These assumptions enable meaningful extrapolation from limited observations.

Neural networks must also incorporate inductive biases to generalize effectively. Although traditional models encode these biases through explicit functional forms, neural networks embed them through architectural design. These architectural choices are not merely computational conveniences but also fundamental ideas about how we can make scientific inferences.

\subsection{Neural Networks as Universal Approximators}
\label{subsection:neural-networks-as-universal-approximators}

To appreciate how modern architectures encode domain knowledge, we begin with feedforward neural networks---also known as multilayer perceptrons or fully connected networks---which make minimal assumptions about data structure \citep{Rosenblatt1958,Rumelhart1986}. These networks represent the simplest form of deep learning: Every neuron in each layer connects to every neuron in the subsequent layer, with information flowing unidirectionally from input to output. This full connectivity allows any input feature to influence any output, making no assumptions about which features might be related. In the language of inductive biases, feedforward networks have minimal bias---they impose almost no structure on the learning problem.

How do these minimally constrained networks achieve their flexibility? Neural networks decompose complex functions into compositions of simple nonlinear transformations. Where Fourier transforms use fixed basis functions (sines and cosines), neural networks learn their basis functions from data. Each neuron computes a weighted sum of its inputs followed by a nonlinear activation, and composing many such operations enables representation of arbitrarily complex relationships.

Mathematically, a feedforward network transforms input $\mathbf{x}$ through $L$ successive layers:
\begin{eqnarray}
\mathbf{h}^{(0)} &=& \mathbf{x} \quad \text{(input layer)},\\
\mathbf{h}^{(l)} &=& f^{(l)}(\mathbf{W}^{(l)}\mathbf{h}^{(l-1)} + \mathbf{b}^{(l)}), \quad l = 1, \ldots, L, \\
\hat{y} &=& \mathbf{h}^{(L)} \quad \text{(output layer)}.
\end{eqnarray}
Here, $\mathbf{h}^{(l)}$ represents activations at layer $l$, $\mathbf{W}^{(l)}$ is the weight matrix connecting layers, and $\mathbf{b}^{(l)}$ is the bias vector. The function $f^{(l)}$ is a nonlinear activation applied elementwise, which is crucial because without it, multiple linear transformations would collapse to a single linear transformation\footnote{Modern deep learning predominantly uses ReLU (rectified linear unit) $f(z) = \max(0, z)$, though alternatives like GELU (Gaussian error linear unit) $f(z) = z \cdot \Phi(z)$ (where $\Phi$ is the cumulative normal distribution) and Swish $f(z) = z \cdot \sigma(z)$ (where $\sigma$ is the sigmoid function) show similar performance. Empirical evidence shows that network performance remains consistent across activation choices that maintain nonvanishing gradients and exhibit sublinear growth \citep{Dubey2021}. For astronomy applications, architectural choices have far greater impact than activation functions.}.

Training finds optimal parameters $\boldsymbol{\theta} = \{\mathbf{W}^{(1)}, \mathbf{b}^{(1)}, \ldots, \mathbf{W}^{(L)}, \mathbf{b}^{(L)}\}$ by minimizing a loss function measuring prediction error. For regression problems predicting continuous values (e.g., stellar parameters), we minimize mean squared error: $\mathcal{L} = (1/N)\sum_{i=1}^{N}(\hat{y}_i - y_i)^2$ from all $N$ training samples. For classification (e.g., galaxy morphology), we use cross-entropy loss. This optimization adjusts millions of parameters simultaneously through gradient descent, which is conceptually similar to maximum likelihood estimation for most classical machine learning techniques but scaled to vastly higher dimensions of parameters.

This flexibility and scalability come from a powerful theoretical foundation. As introduced in Section \ref{subsection:deep-learning-direction}, the universal approximation theorem guarantees that feedforward networks can approximate any continuous function to arbitrary accuracy given sufficient neurons \citep{Cybenko1989,Hornik1989}. Yet this very flexibility reveals that feedforward networks make minimal assumptions---have minimal inductive biases---their minimal structure creates an enormous functional space. Without constraints to guide learning, finding the correct function requires extensive training data. This limitation motivates the architectural innovations we examine next.

\subsection{Encoding Assumptions Through Architecture}
\label{subsection:encoding-assumptions-through-architecture}

Each specialized architecture restricts the functional space by embedding structural assumptions about the data it will encounter---spatial locality in images, temporal ordering in time series, relational structure in graphs, or long-range contextual dependencies in sequences. We now examine four foundational architectures---convolutional neural networks, recurrent neural networks, Transformer architectures, and graph neural networks---with each encoding distinct inductive biases suited to different astronomical data modalities.

\subsubsection{Convolutional neural networks}
\label{subsubsection:convolutional-neural-networks}

Convolutional neural networks (CNNs) constrain the functional space through architectural biases aligned with imaging data, with convolution as the core operation. Rather than connecting every input pixel to every output neuron, CNNs apply learnable filters across spatial locations:
\begin{equation}
y_{ij} = \sum_{m,n} w_{mn} \cdot x_{i+m,j+n} + b,
\end{equation}
where $y_{ij}$ is the output at position $(i,j)$, $w_{mn}$ are filter weights, and $b$ is bias. The filter slides across the image, computing weighted sums of local neighborhoods. 

This approach builds directly on how astronomers have long extracted features from images. Classical methods decompose images using fixed mathematical bases---e.g., Gaussian derivatives for edge detection, Mexican hat wavelets (the second derivative of a Gaussian) for identifying blob-like structures and point sources \citep{Starck1998}. These wavelet transforms provide multiresolution analysis, decomposing images into features at different scales. CNNs take this concept further: Rather than using predetermined filters, CNNs learn optimal feature extractors directly from data. The network discovers which combinations of edges, blobs, and textures best capture the patterns relevant for each specific task.

The key architectural constraint is weight sharing: The same filter kernels apply across all spatial locations. Like fitting low-order polynomials to prevent overfitting, weight sharing restricts models to learn translation-invariant features. The mathematical foundation comes from the convolution theorem---spatial translations become phase shifts that leave Fourier magnitudes unchanged. This ensures the same feature can be detected anywhere in the image, just as a Mexican hat wavelet detects sources regardless of their position.

CNNs build understanding through hierarchical feature learning. Each layer transforms representations from the previous layer, creating a progression from local to global features. Early layers capture edges and textures; intermediate layers detect structures like spiral arms; deeper layers encode entire galaxy morphologies. This works because each layer's receptive field---the input region influencing a neuron---grows with depth, enabling the network to capture patterns at multiple scales simultaneously \citep{Cheng2020}. This multiscale hierarchy parallels the multiresolution analysis of wavelet decompositions but with learned rather than fixed filters.

This hierarchical structure has theoretical foundations in the scattering transform \citep{Bruna2012, Mallat2011}, a mathematical framework that helped researchers understand why CNNs succeed. The scattering transform shows how alternating convolutions with nonlinearities creates cascades that preserve signal properties while building useful invariances. Beyond translation invariance, this architecture provides deformation stability---small input distortions produce only small representation changes. This robustness to variations makes CNNs particularly suited to astronomical images, where objects appear at different orientations, scales, and noise levels.

\begin{marginnote}[]
\entry{Scattering transform}{mathematical decomposition using cascaded wavelet convolutions; provides theoretical foundation for CNNs through alternating convolutions with nonlinearities at multiple scales}
\end{marginnote}

\subsubsection{Recurrent neural networks}
\label{subsubsection:recurrent-neural-networks}

Recurrent neural networks (RNNs) have found application in astronomical time series analysis, though their utility is gradually being superseded by Transformers. RNNs recognize that time series emerge from evolving physical processes. Rather than treating each observation independently, RNNs maintain a ``memory'' that evolves with new observations. Understanding a variable star's current brightness requires knowing not just the instantaneous flux but its pulsation phase---information encoded in the sequence.

Mathematically, an RNN processes a sequence $\{\mathbf{x}_1, \mathbf{x}_2, ..., \mathbf{x}_T\}$ by maintaining a hidden state:
\begin{equation}
\mathbf{h}_t = f(\mathbf{W}_{hh}\mathbf{h}_{t-1} + \mathbf{W}_{xh}\mathbf{x}_t + \mathbf{b}_h),
\end{equation}
where $f$ is a nonlinear activation function, and $\mathbf{W}$ and $\mathbf{b}$ are trainable parameters learned from data. Think of $\mathbf{h}_t$ as the network's notebook at time $t$---it summarizes everything observed so far. At each new observation $\mathbf{x}_t$, the network updates this notebook by combining the previous state $\mathbf{h}_{t-1}$ with the new data. The same update rules apply at every time step, just as physical laws remain constant. To make predictions,
\begin{equation}
\hat{y}_t = g(\mathbf{W}_{hy}\mathbf{h}_t + \mathbf{b}_y),
\end{equation}
where $g$ is an output activation function. The network transforms its current state into a prediction, e.g., the next flux value or the star's variability class.

Hidden Markov models (HMMs) provide the classical analog. Like sorting stars into discrete spectral classes, HMMs assume the system occupies one of a finite number of states---perhaps quiet, rising, or flaring for stellar activity \citep{Stanislavsky2020}. The system transitions between states probabilistically, with each state producing characteristic observations. Although HMMs date to the 1960s \citep{Baum1966}, astronomical adoption awaited high-cadence space missions \citep{Varon2011} and gravitational wave detection \citep{Abbott2017}.

RNNs generalize HMMs by using continuous states. Instead of classifying a star as simply flaring or quiet, RNNs can represent it anywhere in a continuous parameter space. This matches physical reality---stellar oscillations vary smoothly rather than jumping between discrete amplitudes. RNNs encode temporal translation invariance through parameter sharing, mirroring CNNs' spatial approach. Where CNNs apply the same filters across all positions, RNNs apply the same weights at every time step.

However, this parameter sharing also limits memory. Information from early observations must pass through many transformations---like the game of telephone, in which messages degrade with each retelling. Gradients decay exponentially when propagated backward, preventing learning of long-term dependencies \citep{Bengio1994}. Models like long short-term memory (LSTM) \citep{Hochreiter1997} and gated recurrent unit (GRU) \citep{Cho2014} address this through skip connections---direct paths preserving important information. These mechanisms selectively retain crucial information while forgetting noise. This concept also echoes residual connections in CNNs \citep{He2016}, enabling training of networks hundreds of layers deep.

\begin{marginnote}[]
\entry{Residual connections}{shortcuts adding layer input directly to output: $y = f(x) + x$; enables training deeper networks by preserving gradient flow}
\end{marginnote}

Unlike CNNs requiring fixed-size inputs, RNNs naturally handle streaming data by continuously updating their state---ideal for real-time applications in which classifications must evolve as data accumulate. They classify transients \citep{Muthukrishna2019}, predict coronal mass ejections from magnetic fields \citep{Liu2020}, and track galaxy assembly through merger trees \citep{Nguyen2025}. Yet even LSTMs and GRUs compress all history into fixed-size vectors. For surveys with extensive temporal coverage, this compression inevitably loses fine-scale information, motivating the development of Transformers.

\subsubsection{Transformer architectures}
\label{subsubsection:transformer-architectures}

Transformers represent a paradigm shift in sequence modeling, abandoning sequential processing in favor of allowing every observation in a sequence to directly interact with every other observation \citep{Vaswani2017,Dosovitskiy2020}. Transformer architectures are built on the attention mechanism---a mechanism that computes relationships among all pairs of inputs simultaneously.

\begin{marginnote}[]
\entry{Attention mechanism}{neural network operation computing relationships among all input pairs; attention weights determine which input parts influence each output}
\end{marginnote}

CNNs excel when translation invariance helps, but for spectra this becomes a liability. Features at different wavelengths have distinct physical meanings---a line at 5000 \AA{} represents different physics than at 6000 \AA{}. CNNs' sliding kernels inappropriately blend these physically distinct features, performing worse than even simple feedforward networks \citep{Rozanski2025a}. What inductive bias better matches spectral structure? Stellar absorption lines from single elements appear dispersed across wavelengths. Determining abundances requires integrating all these features, weighted by diagnostic power.

The attention mechanism addresses this by computing direct relationships among all wavelengths. It transforms inputs into three learned representations: queries (Q) encoding ``what am I looking for?'', keys (K) encoding ``what do I contain?'', and values (V) encoding ``what information to extract.'' These determine which parts should attend to others:
\begin{equation}
\text{Attention}(\mathbf{Q},\mathbf{K},\mathbf{V}) = \text{softmax}\left(\frac{\mathbf{Q}\mathbf{K}^T}{\sqrt{d_k}}\right)\mathbf{V},
\end{equation}
where $d_k$ is the dimensionality of the key vectors, serving as a normalization factor.

Concretely, given input spectrum $\mathbf{X} = [\mathbf{x}_1, \mathbf{x}_2, ..., \mathbf{x}_n]$ where each $\mathbf{x}_i$ contains flux values at wavelength bin $i$, attention creates these representations through linear transformations:
\begin{align}
\mathbf{Q} &= \mathbf{X}\mathbf{W}_Q \quad \text{(queries: what am I looking for?)}, \\
\mathbf{K} &= \mathbf{X}\mathbf{W}_K \quad \text{(keys: what do I contain?)}, \\
\mathbf{V} &= \mathbf{X}\mathbf{W}_V \quad \text{(values: what information to extract)}.
\end{align}
The weight matrices $\mathbf{W}_Q$, $\mathbf{W}_K$, and $\mathbf{W}_V$ are learnable parameters---analogous to convolution kernels but operating globally rather than locally. This separation allows identifying ``where to look'' (relevant wavelengths), then ``what to extract'' (line strengths and profiles). Through training, these matrices learn task-relevant features.

The matrix product $\mathbf{Q}\mathbf{K}^T$ produces an $n \times n$ attention matrix, where element $(i,j)$ represents how relevant wavelength $j$ is for understanding wavelength $i$. The softmax ensures these relevance scores sum to 1. Finally, multiplying by $\mathbf{V}$ aggregates information---if wavelength $i$ strongly attends to wavelengths containing magnesium lines, the output will emphasize magnesium features. Like stacking convolutional layers in CNNs, Transformers stack attention layers. Each layer's output becomes the next layer's input, building increasingly sophisticated representations. However, unlike CNNs in which deep hierarchies become a liability for long-range dependencies, Transformer layers maintain direct connections across all positions at every level.

This resembles several classical astronomical techniques. Cross-correlation measures radial velocities by comparing observed and template spectra at all wavelength shifts \citep{Simkin1974} and determines galaxy redshifts through similar spectral matching \citep{Tonry1979}. Matched filtering in gravitational wave detection compares strain data against templates at all time lags \citep{Owen1999}. However, these classical methods use fixed similarity metrics, whereas attention learns what features to compare through its trainable weight matrices.

Multihead attention further extends this mechanism by running $h$ parallel attention operations (heads), each with its own set of weight matrices. Each head learns different projection matrices, allowing it to focus on different aspects or relationships in the data. The outputs from all $h$ heads are concatenated and combined through a final projection. This parallels how CNNs learn multiple kernels---though CNNs might learn edge detectors at various orientations, attention heads might specialize in different patterns or relationships. In spectra, where multiple elements appear superimposed, one head might focus on iron lines while another tracks carbon features. The final projection learns how to combine these different views into a unified representation.

Transformers gained prominence in natural language processing by capturing long-range dependencies in text---words separated by entire paragraphs can be semantically related. This capability transfers directly to astronomy: Spectral lines separated by thousands of angstroms are physically related through atomic physics, whereas time series observations separated by months reveal periodic behavior. Indeed, \citet{Rozanski2025a} demonstrated that Transformers outperform both CNNs and feedforward networks on stellar spectra precisely because attention captures these distributed relationships.

Although architectures using multihead attention are often termed Transformers after the original paper \citep{Vaswani2017}, in practice researchers combine different modules for optimal performance. For example, \citet{Pan2024a} uses convolutional layers to extract local features (stellar oscillation modes) and attention layers to model long-range dependencies (granulation patterns affecting mode amplitudes). This hybrid design leverages each module's strengths---convolution efficiently captures the periodic oscillations while attention tracks how granulation modulates these oscillations over longer timescales. This hybrid design leverages each module's strengths: convolution for local periodic features, attention for long-range modulations.

\subsubsection{Graph neural networks}
\label{subsubsection:graph-neural-networks}

Unlike CNNs requiring gridded data or Transformers needing tokenization, graph neural networks (GNNs) naturally handle astronomy's discrete, irregularly distributed objects. Many astronomical observations naturally exist as discrete objects with complex relationships. Galaxy surveys map millions of galaxies as points in 3D space, each with properties like mass, star-formation rate, and morphology. These galaxies do not exist in isolation; they cluster hierarchically, interact gravitationally, and share evolutionary histories through the large-scale environment. Similarly, dark matter halos form merger trees encoding assembly history. GNNs explicitly model such data by representing objects as nodes and relationships as edges \citep{Scarselli2009}.

Astronomers have long used graph-based methods. Minimal spanning trees (MSTs) identify filamentary structure by connecting each galaxy to its nearest neighbors, revealing the large-scale structure \citep{Barrow1985}. The friends-of-friends algorithm groups galaxies by linking those within a threshold distance, essentially finding connected components in a proximity graph \citep{Turner1976}. Related methods use local galaxy density for photometric redshift estimation, though without explicit graph construction \citep{Menard2013}. These classical approaches use fixed rules for both connectivity and analysis. An MST connects galaxies by minimum distance; friends-of-friends links by a constant threshold. They do not learn how properties propagate.

GNNs build on these graph-based ideas but learn the relationships from data rather than imposing fixed rules. A GNN operates on a graph $\mathcal{G} = (\mathcal{V}, \mathcal{E})$, where nodes $\mathcal{V}$ represent objects, and edges $\mathcal{E}$ encode relationships. Each galaxy node $i$ has features $\mathbf{h}_i$ (mass, gas fraction, star-formation rate), whereas edges carry attributes $\mathbf{e}_{ij}$ (spatial distance, velocity).

The core operation is message passing: Nodes update their representations by aggregating neighbor information across layers:
\begin{equation}
\mathbf{h}_i^{(l+1)} = \phi^{(l)}\left(\mathbf{h}_i^{(l)}, \bigoplus_{j \in \mathcal{N}(i)} \psi^{(l)}(\mathbf{h}_i^{(l)}, \mathbf{h}_j^{(l)}, \mathbf{e}_{ij})\right)
\end{equation}
where superscript $(l)$ denotes the layer index. The aggregation operator $\bigoplus$ (sum, mean, or max) ensures permutation invariance; reordering neighbors does not change the result, just as the mean mass of galaxies does not depend on their catalog order. Think of this as galaxies communicating with their neighbors. The message function $\psi$ determines what information galaxy $j$ sends to galaxy $i$---perhaps how its star formation might influence neighbors. The aggregation $\bigoplus$ combines messages from all neighbors $\mathcal{N}(i)$. The function $\phi$ then updates galaxy $i$'s representation based on its current state and received messages.

The functions $\psi^{(l)}$ and $\phi^{(l)}$ are neural networks with learnable parameters---these are optimized during training. Given input node features (observed galaxy properties) and a training objective (e.g., predicting star-formation rates), the GNN learns what information to pass between neighbors and how to update representations. The goal is to find message-passing rules that optimize the final predictions. After $L$ layers, each node incorporates information from objects up to $L$ connections away. The final representations $\mathbf{h}_i^{(L)}$ enable node-level predictions (individual galaxy properties) or graph-level predictions (entire cluster properties).

Although standard GNNs aggregate information through explicit edges, some architectures modify this approach. PointNet \citep{Qi2016}, for instance, can be viewed as a GNN in which the graph is fully connected---every node connects to every other node. This architecture proves particularly effective for point cloud data like galaxy distributions, where the ordering of objects is arbitrary but their collective properties matter.

This layered propagation---whether through sparse edges or fully connected architectures---mirrors physical processes: Gravitational influences cascade through structure; chemical enrichment spreads between interacting systems. The learned message passing can discover which relationships matter most---perhaps finding that galaxy properties depend more on dark matter environment than simple spatial proximity.

Applications demonstrate GNNs' power for relational data. For galaxy intrinsic alignments, a key systematic in weak lensing, \citet{Craigie2025b} showed GNNs learn alignments from local galaxy environments in observational data, whereas \citet{Jagvaral2022} connected alignments to physical properties like tidal fields and halo mass using simulations. GNNs excel at reconstruction tasks with irregular graph structure. \citet{Tang2022} reconstructed galaxy merger histories, where progenitor systems form trees with variable numbers of nodes and relationships. Similarly, \citet{Jespersen2022} predicted baryon properties from dark matter merger trees; the GNN learns how baryon physics depends on assembly history encoded in the merger graph.

Recent work demonstrates that GNNs can perform cosmological parameter inference directly from discrete galaxy catalogs. \citet{Villanueva-Domingo2022} and \citet{Lee2024} treat galaxy positions as graph nodes. These GNNs learn which spatial scales and relationships constrain cosmological parameters, potentially discovering informative features beyond traditional two-point statistics.

\begin{table}[h]
\tabcolsep5pt
\caption{Architectural choices as encoded physical hypotheses}
\label{table1}
\begin{center}
\begin{tabular}{@{}l|l|l|l@{}}
\hline
\textbf{Architecture} &\textbf{Inductive bias} &\textbf{Classical equivalent} &\textbf{When to use}\\
\hline
Convolutional &Translation invariance &Wavelet analysis &Images with hierarchical,\\[-2pt]
 neural network &Hierarchical features & &translation-invariant\\[-2pt]
 &Deformation stability & &features\\[3pt]
\hline
Recurrent neural &Temporal invariance &Hidden Markov models &Short-to-medium sequences\\[-2pt]
 network &Sequential memory &Autoregressive models &with temporal\\[-2pt]
 &Causal ordering & &dependencies and\\[-2pt]
 & & &streaming data\\[3pt]
\hline
Transformer &Global connectivity &Cross-correlation &Long-range dependencies in\\[-2pt]
 &Learned attention &Template matching &spectra and time series\\[3pt]
\hline
Graph neural &Permutation invariance &Minimal spanning trees &Discrete objects with\\[-2pt]
 network &Local aggregation &Friends-of-friends &relational structure\\
\hline
\end{tabular}
\vspace{-0.8cm}
\end{center}
\end{table}

\subsubsection{On architectural choices}
\label{subsubsection:on-architectural-choices}

The architectures presented demonstrate that effective models must align their inductive biases with the structure of astronomical data. Each specialized architecture reduces this search space by encoding symmetries (see Section~\ref{subsection:encoding-physical-symmetries}) and relationships inherent in different data types. \textbf{Table~\ref{table1}} summarizes these architectural choices and their astronomical applications.

These connections reveal that deep learning techniques build on established astronomical methods by making them learnable. Where wavelet transforms use fixed basis functions, CNNs learn optimal ones from data. Where HMMs require discrete states, RNNs learn continuous representations. Where cross-correlation uses rigid templates, Transformers learn flexible attention patterns. The innovation lies not in abandoning classical methods but in making them adaptive. When architectural assumptions align with data structure, models achieve better performance with far less training data. Conversely, architectural mismatch forces models to compensate with unnecessarily large datasets or leads to convergence failure. Success depends on understanding your data's structure and choosing architectures whose inductive biases match that data structure.

\subsection{Encoding Physical Symmetries}
\label{subsection:encoding-physical-symmetries}
    
The architectural biases examined previously---translation invariance in CNNs, temporal invariance in RNNs, and permutation invariance in GNNs---represent specific instances of a broader principle: encoding physical symmetries into neural networks. In physics, symmetries are transformations that leave properties unchanged. Conservation laws emerge from symmetries: energy conservation from time translation symmetry, momentum conservation from spatial translation symmetry. Similarly, neural networks can encode these symmetries, ensuring outputs respect the same physical principles. Rather than hoping networks discover these symmetries from data alone, we can build them directly into the architecture.

Consider classifying galaxy morphologies. The network needs detailed spatial information---where spiral arms appear, how bars orient, where bulges sit---to distinguish different types. Yet the classification should not depend on where the galaxy appears in the image or how it's rotated. The network must preserve structural information during processing but produce position-independent answers.

This requirement motivates two concepts. Let $g$ denote a transformation acting on input $x$---such as shifting an image, rotating it, or reordering objects in a list. A function $f$ is \textit{invariant} if its output remains unchanged under transformation,
\begin{equation}
f(g[x]) = f(x).
\end{equation}
For example, the total flux from a galaxy is invariant to rotation; i.e., spinning the galaxy image does not change the integrated brightness. A function is \textit{equivariant} if its output transforms in the same way as the input,
\begin{equation}
f(g[x]) = g[f(x)].
\end{equation}
If we rotate a galaxy image by 90°, a feature detector for spiral arms should also rotate its output by 90°. The features move with the galaxy.

\begin{marginnote}[]
\entry{Invariant}{function output unchanged by input transformation}
\entry{Equivariant}{function output transforms consistently with input transformation; describes how networks handle symmetries}
\end{marginnote}

The blueprint for encoding symmetries combines equivariant feature extraction with invariant final predictions. Intermediate layers maintain equivariance to preserve geometric structure, whereas the final layer computes invariant outputs through operations like global averaging \citep{Cohen2016,Bronstein2021}. This maintain-structure-then-summarize principle is crucial: Though direct invariant operations correctly produce identical outputs for transformed inputs, they lack expressiveness by immediately discarding spatial structure. In contrast, equivariant layers preserve geometric relationships---rotated galaxies produce correspondingly rotated feature maps---before a final invariant aggregation yields consistent classifications (see \textbf{Supplemental Text, Section C} for a visual illustration).

Standard CNNs highlight this principle. Convolution is translation equivariant---shifting the input shifts all feature maps equally:
\begin{equation}
f_{\text{conv}}(\text{shift}_{dx,dy}[x]) = \text{shift}_{dx,dy}[f_{\text{conv}}(x)].
\end{equation}
Through multiple convolutional layers, CNNs build hierarchical features while preserving their spatial relationships. Only at the final layer does global pooling achieve translation invariance for classification.

This principle---maintain structure then summarize---unifies modern architectures. RNNs exhibit temporal equivariance through shared weights. If $\text{shift}_\tau$ denotes shifting a sequence by $\tau$ time steps, then
\begin{equation}
f_{\text{RNN}}(\text{shift}_\tau[\mathbf{x}]) = \text{shift}_\tau[f_{\text{RNN}}(\mathbf{x})].
\end{equation}
The same transformation $f(\mathbf{W}_{hh}\mathbf{h}_{t-1} + \mathbf{W}_{xh}\mathbf{x}_t + \mathbf{b}_h)$ applies at every time step, ensuring that delaying a light curve by one day delays all extracted features by one day. This preserves temporal structure until final aggregation yields time-invariant predictions. GNNs maintain permutation equivariance through symmetric operations. If $\pi$ denotes a permutation of nodes, then
\begin{equation}
f_{\text{GNN}}(\pi[\mathbf{X}]) = \pi[f_{\text{GNN}}(\mathbf{X})],
\end{equation}
where $\mathbf{X}$ represents all node features. The aggregation operator ensures that reordering galaxies in a catalog reorders their output representations identically. This preserves relational structure---galaxy $i$'s representation always depends on its neighbors, regardless of catalog ordering---until invariant pooling computes cluster properties. Nonetheless, standard architectures encode limited symmetries. Astronomical data often require richer symmetries, which have inspired other physics-inspired neural networks.

\subsubsection{Rotation invariance}
\label{subsubsection:rotation-invariance}
Many astronomical observations lack preferred orientations. Galaxies orient randomly on the sky, and the cosmic microwave background shows no special directions. Yet standard CNNs fail under rotation; rotating a galaxy image produces unpredictably transformed features rather than correspondingly rotated ones. CNNs must learn each rotated version of objects separately, wasting parameters and training data.

Rotation-equivariant convolutions address this by extending the equivariance principle beyond translations. Instead of convolving only over spatial positions $(x,y)$, they convolve over positions and orientations $(x,y,\theta)$. Each layer maintains rotation equivariance: Rotating the input galaxy rotates all feature maps correspondingly. The network learns one set of filters that automatically work at all orientations, preserving angular structure through multiple layers. Only the final aggregation over orientations yields rotation-invariant classifications.

For galaxy morphology classification, rotation-equivariant networks \citep{Weiler2019} have shown improved robustness to corrupted pixels, reduced rotational uncertainty, and more meaningful latent representations. By encoding the correct inductive bias, these networks achieve better data efficiency---learning from fewer examples---while generalizing more reliably and remaining less susceptible to artifacts or noise \citep{Scaife2021,Pandya2023}. These symmetry principles extend beyond individual objects. Encoding these symmetries enables more efficient learning for large-scale structures \citep{Dai2022}.

\begin{marginnote}[]
\entry{Latent representation}{lower-dimensional representation capturing essential features of high-dimensional data; similar objects cluster together, enabling interpolation and relationship discovery}
\end{marginnote}

\subsubsection{Scale invariance}
\label{subsubsection:scale-invariance}

Power laws in astronomy signal scale invariance. A distribution $P(x) \propto x^{-\alpha}$ maintains its functional form under rescaling: $P(\lambda x) = \lambda^{-\alpha} P(x)$. This means the same physics governs phenomena across different scales. This property appears throughout astrophysics: Kolmogorov turbulence produces velocity-size-mass relations in molecular clouds \citep{Larson1981}, and the halo mass function follows power laws in intermediate-mass regimes \citep{Press1974}.

To encode scale invariance, we need architectures in which rescaling the input produces predictable transformations of features. The insight is applying the same convolutional filters at logarithmically spaced scales. If we have a filter $w$ that detects features at scale $\lambda$, we apply rescaled versions at scales $2\lambda$, $4\lambda$, $8\lambda$, and so on:
\begin{equation}
y_{ij}^{(s)} = \sum_{m,n} w_{mn}^{(s)} \cdot x_{i+m,j+n},
\end{equation}
where $w^{(s)}$ is the same pattern rescaled to size $2^s$. This ensures that zooming the input by a factor of 2 simply shifts features between scale channels. Consider detecting a turbulent eddy. If we zoom in by a factor of 2, the eddy appears twice as large. With scale-equivariant convolutions, this eddy shifts from the $s$-th to the $(s+1)$-th scale channel, enabling recognition regardless of scale.

The scattering transform implements this principle through a cascade of wavelet convolutions and modulus operations \citep{Bruna2012,Mallat2011}. Scattering transform was originally developed to understand deep network success. It uses fixed mathematical operations: convolving with wavelets at scales $\lambda_s = 2^{s}$, taking the modulus for stability, then iterating. Each iteration captures interactions between different scales, building a multiscale representation without learning---like a fixed, mathematically designed CNN. The success of the scattering transform in cosmology \citep{Cheng2020,Valogiannis2022} and turbulence studies \citep{Allys2019} stems from this alignment. Scale-equivariant neural networks extend this by making filters learnable while preserving the scale structure. \citet{Craigie2025a} demonstrated this approach's power by detecting parity violation in simulated cosmological data, where the scale-equivariant architecture proved more sample-efficient than standard CNNs.

\subsubsection{Lorentz invariance}
\label{subsubsection:lorentz-invariance}

Relativistic astrophysics demands Lorentz symmetry: invariance under coordinate transformations mixing space and time. Relativistic jets exhibit extreme Lorentz factors. Gamma-ray bursts, cosmic ray acceleration, and gravitational waves all involve relativistic effects: time dilation, length contraction, and beaming. Particle physicists have developed Lorentz-equivariant networks encoding special relativity into their structure \citep{Bogatskiy2020,Gong2022}. These architectures ensure proper transformation of four-vectors (e.g., spacetime position or energy-momentum) and preservation of relativistic invariants. Networks automatically respect relativity rather than learning it from data. Deep learning has achieved notable success in relativistic astrophysics, with applications to gravitational wave detection \citep{George2018,Dax2021} and high-energy phenomena \citep{Erdmann2018}. However, the potential of Lorentz-equivariant architectures remains unexplored in astronomy, making this an interesting direction for future research.

\subsubsection{From physical symmetries to architectural design}
\label{subsubsection:from-physical-symmetries-to-architectural-design}

The incorporation of physical symmetries into neural network architectures represents a systematic approach to encoding domain knowledge. Although data augmentation---artificially expanding training sets by applying transformations like rotations or shifts to existing data---can help networks learn approximate symmetries, architectures with built-in equivariance guarantee exact symmetry preservation while requiring fewer parameters and less training data. This distinction matters for astronomical applications in which labeled datasets are often limited. Each symmetry constrains the function space networks explore. By limiting search to physically plausible solutions, these constraints improve model performance while making neural networks more data efficient. These physics-inspired architectures demonstrate that astronomy and machine learning advance together. While AI accelerates astronomical discovery, physical insights from astronomy push machine learning forward. This bidirectional flow---encoding physics into architectures that then enable new discoveries---illustrates how domain knowledge in astrophysics and computational methods reinforce each other.



\subsection{Conservation Laws and Equations as Constraints}
\label{subsection:conservation-laws-and-equations-as-constraints}

Beyond symmetries, astrophysical systems obey conservation laws and differential equations that provide additional structure for neural network training. Solutions satisfying the underlying equations generalize better; they can extrapolate to conditions not in the training data. Although purely data-driven models might memorize specific examples, physics-informed models learn the underlying dynamics governing the system everywhere.

Neural networks are differentiable by design---this is how we train them through gradient descent. This same property makes them naturally suited for incorporating differential equations as constraints. Because ordinary differential equations (ODEs) and partial differential equations are ubiquitous in physics---from orbital mechanics to fluid dynamics---we can leverage the network's differentiability to ensure solutions respect these equations. Modern frameworks compute these derivatives automatically and exactly, making it straightforward to incorporate equations of arbitrary complexity.

The key idea is to also train networks on the governing physics. In standard supervised learning, we minimize a loss function that measures prediction error. Physics-informed approaches expand this by adding terms that penalize violations of physical laws,
\begin{equation}
\mathcal{L} = \mathcal{L}_\text{data} + \lambda_\text{physics} \mathcal{L}_\text{physics}.
\end{equation}
In this architecture, a neural network $f_\theta$ processes input coordinates $\mathbf{x}$ to produce both function values $f(\mathbf{x})$ and derivatives $\partial f/\partial \mathbf{x}$ through automatic differentiation. The data loss $\mathcal{L}_\text{data}$ measures agreement between predicted and observed values, whereas the physics loss $\mathcal{L}_\text{physics}$ penalizes violations of governing equations. This dual constraint enables networks to learn solutions that simultaneously fit observations and respect physical laws (see \textbf{Supplemental Text, Section C}).

When incorporating differential equations, neural networks serve as continuous representations of physical fields. The physics loss measures how well this representation satisfies the governing equations by evaluating equation residuals at sampled points throughout the domain. The weight $\lambda_\text{physics}$ balances these objectives: Larger values more strongly enforce the corresponding constraints. This forces the network to find solutions that simultaneously fit the data and obey the physics. Viewed another way, this approach finds solutions to differential equations without assuming specific functional forms. Traditional analytical methods often require guessing solution structures (like assuming separability in frequency space), whereas these physics-informed approaches learn flexible mappings that satisfy the constraints. The network discovers the appropriate functional form through optimization.

\subsubsection{Illustrative example: nonparametric galactic dynamics}
\label{subsubsection:illustrative-example-non-parametric-galactic-dynamics}

This approach of incorporating differential equations as constraints has proven particularly effective for inferring the Milky Way's gravitational potential from stellar kinematics, which is a classic inverse problem in which we observe consequences (stellar motions) but seek causes (gravitational potential).

Recovering the gravitational potential $\Phi(\mathbf{x})$ from stellar observations remains challenging because we measure only positions $\mathbf{x}$ and velocities $\mathbf{v}$, not the underlying potential. The key physical constraint is that in steady state, the phase-space distribution function $f(\mathbf{x}, \mathbf{v})$ must satisfy the collisionless Boltzmann equation,
\begin{equation}
\frac{df}{dt} = \sum_i \left( v_i \frac{\partial f}{\partial x_i} - \frac{\partial \Phi}{\partial x_i} \frac{\partial f}{\partial v_i} \right) = 0.
\end{equation}
This equation encodes stellar dynamics: The term $v_i (\partial f/\partial x_i)$ represents how the distribution evolves as stars move through space, whereas $(\partial \Phi/\partial x_i) (\partial f/\partial v_i)$ captures how gravitational forces accelerate stars and reshape velocities. At equilibrium, these effects balance exactly, linking observable stellar distributions to the invisible gravitational potential.

Traditional approaches assume specific functional forms---isothermal spheres, Plummer models, or King models \citep{Binney2008}---that yield tractable relationships between $f$ and $\Phi$. However, such parametric assumptions may not reflect reality. The choice of model can introduce systematic biases in recovered mass distributions \citep{deBlok2010}.

\citet{Green2020a} demonstrates how neural networks solve this problem without restrictive assumptions \citep[see also][]{An2021,Buckley2023}. The approach combines data and physics constraints. First, a neural network $f_\phi$ (where $\phi$ denotes the network parameters) learns the phase-space distribution from observed stellar kinematics:
\begin{equation}
\mathcal{L}_\text{data} = -\sum_{\text{stars}} \ln f_\phi(\mathbf{x}_k, \mathbf{v}_k),
\end{equation}
where $(\mathbf{x}_k, \mathbf{v}_k)$ are the observed positions and velocities of individual stars. The gravitational potential is parameterized as a neural network $\Phi_\theta(\mathbf{x})$ with parameters $\theta$ and trained to satisfy physical constraints. The physics loss enforces that the distribution remains stationary under the learned potential:
\begin{equation}
\mathcal{L}_\text{physics} = \left\langle \left| \sum_i \left( v_i \frac{\partial f}{\partial x_i} - \frac{\partial \Phi_\theta}{\partial x_i} \frac{\partial f}{\partial v_i} \right) \right|^2 \right\rangle_{\mathbf{x},\mathbf{v} \sim f},
\end{equation}
where $i$ indexes the spatial dimensions, and the angle brackets $\langle \cdot \rangle$ denote averaging over phase-space points sampled from the distribution $f$.

These deep learning approaches that solve equations while maintaining physical consistency open new opportunities in galactic dynamics without restrictive assumptions. \citet{Kalda2025} analyzed Gaia DR3 stars, recovering local pattern speed and matter density without parametric assumptions. Complementary work by \citet{Lim2025} measured the local dark matter density using deep learning surrogate models.

\subsubsection{Other applications}
\label{subsubsection:other-applications}

Similar physics-informed approaches have been applied across diverse astrophysical domains. In stellar astrophysics, \citet{Li2025} enforced hydrostatic equilibrium in stellar atmosphere models. In solar physics, \citet{Baty2024} incorporated magnetohydrodynamic principles: force-free conditions for magnetic arcades, Grad-Shafranov equilibrium for curved structures, and resistive magnetohydrodynamics equations for reconnection. For astrophysical shocks, \citet{Moschou2023} used conservation laws to model solar wind from corona to termination shock. Enforcing mass, momentum, and energy conservation ensured physically consistent solutions across the wind's acceleration, heating, and shocking. In interstellar medium (ISM) chemistry, \citet{Branca2023} trained networks to satisfy reaction rate equations for chemical species along with energy balance.

\citet{Greydanus2019} demonstrated that neural networks can learn the Hamiltonian itself. By parameterizing $H(q,p)$ as a neural network and computing dynamics through Hamilton's equations, the network learns from observed trajectories while guaranteeing energy conservation \citep[see also Lagrangian neural networks,][]{Cranmer2020b}, which parameterize the Lagrangian $L(q,\dot{q})$ instead. However, realistic astrophysical applications of such Hamiltonian neural networks and Lagrangian neural networks with great dynamic ranges remain challenging, often requiring hybrid approaches that combine neural predictions with numerical integration \citep{SazUlibarrena2024}.

\subsection{Balancing Physical Constraints with Discovery}
\label{subsection:balancing-physical-constraints-with-discovery}
    
The preceding examples illustrate that deep learning provides new ways to encode our physical understanding. Physics-informed approaches---through symmetry-aware architectures or conservation law constraints---represent powerful inductive biases that enable generalization from limited astronomical data. Yet the choice of constraints matters. For example, enforcing the collisionless Boltzmann equation in galactic dynamics assumes equilibrium, but real galaxies may violate steady-state assumptions. This tension between constraint and discovery pervades all astronomical modeling. The advantage of modern deep learning is flexibility: We can encode symmetries through equivariant operations, enforce conservation laws through differentiable constraints, and relax these when data demand it. The challenge lies in choosing which physical principles to encode as hard architectural constraints, which to impose as soft regularization, and which to leave flexible for discovery. In practice, loss functions take the form $\mathcal{L} = \mathcal{L}_\text{data} + \lambda_\text{physics} \mathcal{L}_\text{physics}$, where hyperparameters like $\lambda_\text{physics}$ control how strongly we enforce theoretical priors.

Physics-informed priors can guide models toward physically meaningful solutions without sacrificing flexibility. For instance, theoretical models correctly identify which spectral features trace which elements, even when synthetic spectra contain systematic errors. By constraining neural networks to measure abundances through their actual atomic and molecular features rather than element-element correlations, \citet{Ting2017} and \citet{Xiang2019} ensured physically meaningful stellar chemical measurements. The most successful approaches balance physical knowledge with data-driven flexibility. This synthesis represents not a compromise but an evolution in how we encode and test theoretical knowledge against observations.

\section{CROSS-CUTTING TECHNIQUES}
\label{section:cross-cutting-techniques}
    
The building blocks from Section~\ref{section:methodological-foundations}—architectures (Section~\ref{subsection:encoding-assumptions-through-architecture}), symmetry constraints (Section~\ref{subsection:encoding-physical-symmetries}), and physics-informed losses (Section~\ref{subsection:conservation-laws-and-equations-as-constraints})—combine flexibly for astronomical problems.  The modular nature of these components enables us to construct specialized solutions by mixing architectures, loss terms, and constraints appropriate to their data and physics. The most impactful applications emerge when these building blocks combine to address challenges that span multiple scales, data modalities, and physical regimes. This section examines techniques that transcend traditional subfield boundaries.

\subsection{Multiscale Modeling and Simulation Surrogates}
\label{subsection:multi-scale-modeling-and-simulation-surrogates}

Astronomical systems span vast ranges of spatial and temporal scales. Galaxy formation involves processes from stellar scales to large-scale environments, which is a dynamic range exceeding ten orders of magnitude. Direct numerical simulation across all scales remains computationally prohibitive. Simulations face the tradeoff of resolution versus volume. For instance, galaxy-formation simulations typically achieve parsec-scale resolution in small volumes or kiloparsec resolution in cosmological volumes, but not both. Processes below the resolution limit---individual supernovae, stellar winds, or molecular cloud collapse---must be approximated through subgrid models.

These subgrid models use parametric prescriptions to capture unresolved physics. In galaxy simulations, they describe how gas within a resolution element forms stars and how stellar feedback heats surrounding gas. Similarly, cosmological N-body simulations efficiently model dark matter dynamics but require additional modeling to connect dark matter halos to observable galaxies. Parametric models make simplifying assumptions: treating star formation as dependent only on local gas density, ignoring multiphase ISM structure, or using averaged feedback efficiencies that miss stochastic stellar populations \citep{Naab2017}.

\subsubsection{Learning multiscale mappings}
\label{subsubsection:learning-multi-scale-mappings}

Deep learning offers a data-driven alternative to parametric subgrid models: Learn effective theories directly from high-resolution simulations. As successful parametric models already demonstrate that low-dimensional descriptions can capture essential physics, this motivates using neural networks to discover these compressed representations automatically. Consider baryonification---adding realistic galaxies to dark matter-only simulations. Traditional approaches use halo occupation statistics and empirical scaling relations. Neural networks can instead learn the complex dark matter-to-galaxy mapping directly from hydrodynamic simulations. The challenge is encoding the right inductive biases to ensure generalization rather than memorization.

\begin{marginnote}[]
\entry{Encoder--decoder}{architecture compressing inputs to compact representations (encoder), then reconstructing outputs (decoder); bottleneck forces discovery of essential features}
\end{marginnote}

Encoder-decoder architectures provide this structure. An encoder network compresses high-dimensional input $\mathbf{x}$ (e.g., a dark matter density field) to a compact latent representation $\mathbf{z}$ (a low-dimensional vector encoding essential features):
\begin{equation}
\mathbf{z} = f_{\text{encoder}}(\mathbf{x}).
\end{equation}
This bottleneck---compressing high-dimensional pixel fields to vectors of tens to hundreds of numbers---forces the network to identify essential features. A decoder then reconstructs the target $\mathbf{y}$ (e.g., galaxy distribution) from this compressed encoding:
\begin{equation}
\mathbf{y} = f_{\text{decoder}}(\mathbf{z}).
\end{equation}
Unlike hand-crafted summary statistics, these learned latent representations aim to capture complex correlations: effective halo occupation, environment-gas relationships, and nonlinear property combinations. The compression mitigates overfitting while the learned features adapt to the specific physics being modeled.

Particularly, U-Net \citep{Ronneberger2015} has gained prominence in astronomical research. This method enhances the encoder-decoder approach by adding skip connections---pathways that bypass the compression bottleneck---between encoder and decoder layers. These connections preserve fine-scale spatial information (like individual galaxy positions) while the compressed pathway captures global context (like large-scale environment), which is crucial for reconstructing small-scale galactic features within large-scale structure.

\subsubsection{Illustrative example: baryonification of N-body simulations}
\label{subsubsection:illustrative-example-baryonification-of-n-body-simulations}

\begin{figure}[ht!]
\centering
\includegraphics[width=\textwidth]{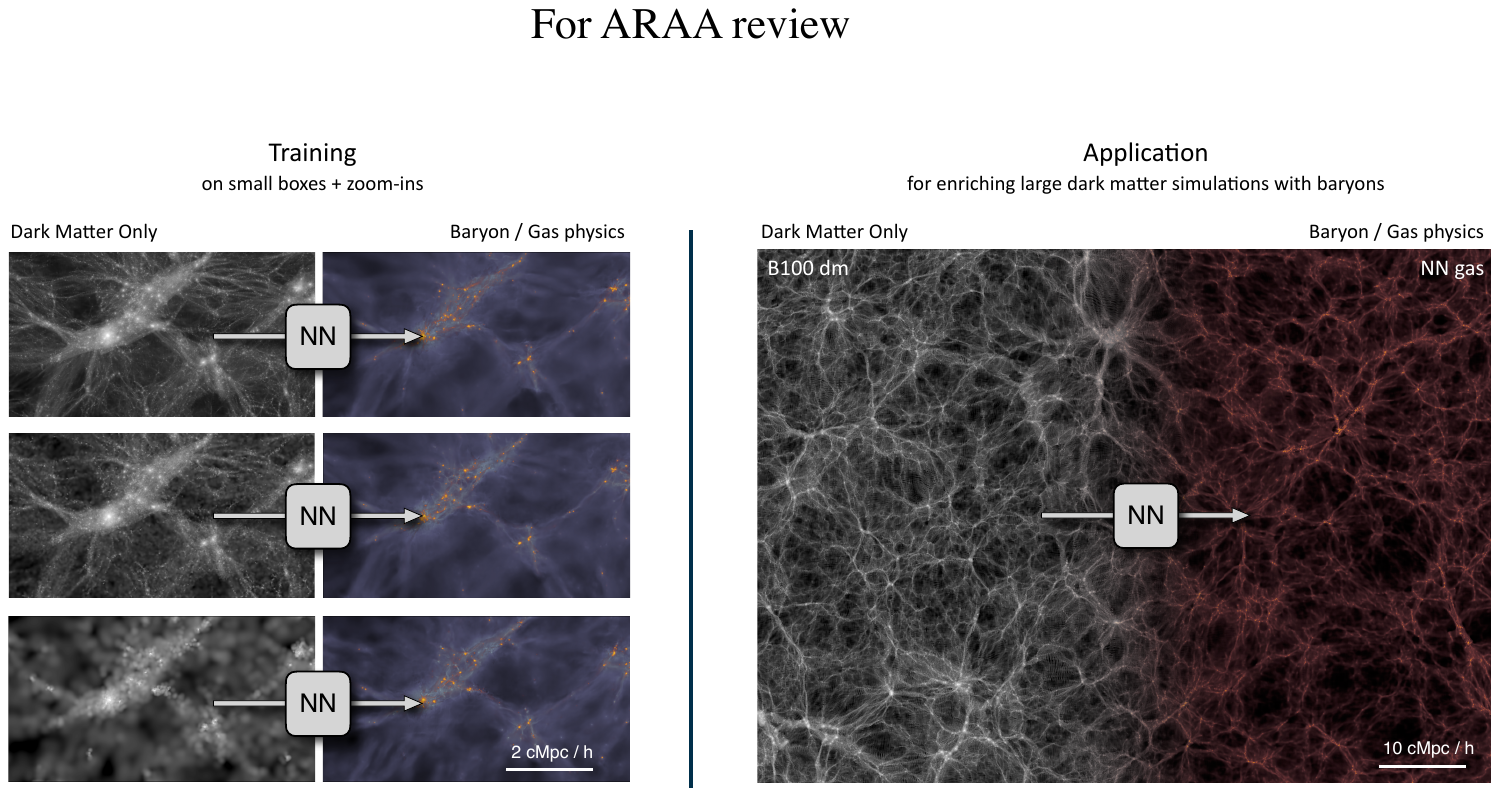}
\caption{Deep learning for multiscale modeling in astrophysics. (\textit{a}) NNs learn mappings between scales by training on paired low- and high-resolution simulations. (\textit{b}) Trained models act as learned subgrid prescriptions, adding small-scale physics to large-volume simulations. This deep learning approach learns directly from high-fidelity simulations rather than using fixed parametric models. Figure adapted with permission from \citet{Bernardini2022}. Abbreviation: NN, neural network.}
\label{fig2}
\end{figure}

U-Net architectures excel at the baryonification problem introduced earlier. Networks learn mappings from dark matter density to multiple gas properties,
\begin{equation}
f_\theta: \rho_\text{DM}(\mathbf{x}) \rightarrow \{\rho_\text{gas}(\mathbf{x}), \rho_\text{HI}(\mathbf{x})\},
\end{equation}
where $\rho_\text{DM}$, $\rho_\text{gas}$, and $\rho_\text{HI}$ denote dark matter, gas, and neutral hydrogen densities, respectively, at spatial position $\mathbf{x}$, with $\theta$ representing the network parameters. Training on hydrodynamical simulations, these surrogates add realistic baryon physics to computationally cheaper dark matter simulations \citep[][see \textbf{Figure~\ref{fig2}}]{Bernardini2022}. The network takes projected dark matter density fields as input and predicts corresponding gas density and neutral hydrogen distributions. Once trained, this surrogate can ``paint'' gas physics onto dark matter--only simulations that would be computationally prohibitive to run.

Applications of such multiscale mapping extend across astrophysical scales. For galaxy clusters, \citet{Chadayammuri2023} train U-Nets to predict X-ray observables from dark matter mass maps. Their network learns the complex relationship between dark matter and hot gas emission, predicting projected gas density, temperature, and X-ray surface brightness. At galactic scales, \citet{Hirashima2025} employ similar architectures to accelerate galaxy simulations by learning pre- to postsupernova gas state mappings. Rather than explicitly simulating each explosion, their network learns how supernova feedback transforms the surrounding gas.

\begin{marginnote}[]
\entry{Simulation-based inference (SBI)}{Bayesian inference when likelihoods are intractable but simulations feasible; neural networks approximate posteriors or likelihoods from simulations}
\end{marginnote}

Beyond deterministic U-Nets, astronomical research in recent years has seen other architectures for capturing stochastic small-scale physics. Diffusion models and normalizing flows---discussed further in the simulation-based inference (SBI) section---can learn the full probability distribution of small-scale properties given large-scale conditions. \citet{Xu2023} demonstrate this by using diffusion models to infer 3D molecular cloud density structure from 2D column density observations.

We note that these surrogate models offer computational advantages but with important caveats. Neural networks trained on specific simulations inherit their systematic biases. Different subgrid implementations can produce similar large-scale observables while yielding divergent predictions for small-scale structure \citep{Li2020}. If the training simulations have unresolved systematics, the surrogate might faithfully reproduce these errors.

\subsubsection{Learning dynamical evolution with neural ordinary differential equations}
\label{subsubsection:learning-dynamical-evolution-with-neural-odes}

The U-Net approaches discussed above learn static mappings between scales; e.g., given dark matter density, predict gas density. Although effective for specific tasks, these mappings are inherently limited: They capture correlations in training data but may not generalize to different physical regimes or simulation parameters. A more principled approach recognizes that astrophysical systems evolve through differential equations. Rather than learning static mappings, we can learn the dynamical equations themselves---incorporating stronger inductive bias through the mathematical structure of evolution.

Neural ODEs provide this framework \citep{Chen2018}. They parameterize the time evolution of a system state $\mathbf{z}$ through a neural network,
\begin{equation}
\frac{d\mathbf{z}}{dt} = f_\theta(\mathbf{z}(t), t),
\end{equation}
where $f_\theta$ learns the governing dynamics with parameters $\theta$. Neural ODEs make the solution process differentiable, enabling gradient-based learning. Given trajectories from high-resolution simulations, we train the network to minimize the difference between predicted and true evolution.

Consider dynamical friction as an example. When a massive object moves through lighter particles, it experiences drag from gravitational wake effects. Chandrasekhar derived the leading-order term analytically,
\begin{equation}
\frac{d\mathbf{v}}{dt} = -\eta(M, \mathbf{v}, \rho, \sigma) \mathbf{v},
\end{equation}
where the friction coefficient $\eta$ depends on object mass $M$, velocity $\mathbf{v}$, background density $\rho$, and velocity dispersion $\sigma$ \citep{Chandrasekhar1943}.

However, analytical derivations capture only leading-order effects. Real systems exhibit corrections from inhomogeneities, anisotropies, and nonequilibrium effects that defy analytical treatment. This motivates hybrid formulations,
\begin{equation}
\frac{d\mathbf{z}}{dt} = f_\text{known}(\mathbf{z}(t), t) + f_\theta(\mathbf{z}(t), t),
\end{equation}
where $f_\text{known}$ encodes established physics (gravity, hydrodynamics) while the neural network $f_\theta$ learns residual corrections. This decomposition ensures the model respects known conservation laws while learning unknown physics from data.

\subsubsection{Illustrative example: galactic winds and cloud evolution}
\label{subsubsection:illustrative-example-galactic-winds-and-cloud-evolution}

Galactic gas dynamics is an area where applying neural ODEs excel because they are systems governed by known hydrodynamics but shaped by poorly understood multiphase interactions. Two recent applications demonstrate how embedding neural networks within differential equations can discover these hidden physics.

\citet{Nguyen2023} tackled mass-loading in galactic winds from starburst galaxy M82. Mass-loading---how cool clouds are destroyed and incorporated into hot winds---involves radiative turbulent mixing, cloud crushing, and multiphase interactions that depend sensitively on local conditions. Rather than guessing functional forms, they parameterized the mass-loading rate as a neural network within the wind evolution equations. The approach discovered nontrivial structure without prior assumptions, revealing sharp truncation at specific radii that analytical models miss.

\citet{Tan2025} applied similar methods to infalling clouds in galactic environments, where cold gas clouds interact with hot ambient media through turbulent mixing layers. They embedded neural networks within differential equations to learn the weight factor governing Kelvin-Helmholtz instabilities and the drag coefficient. Training on high-resolution 3D simulations, the neural ODE framework learned corrections that better captured cloud evolution than analytical models alone. Both studies exemplify how neural ODEs extract hidden physics when underlying processes defy analytical treatment, discovering functional forms directly from data while respecting established conservation laws.

\subsubsection{A cautionary note: super-resolution misconceptions}
\label{subsubsection:a-cautionary-note-super-resolution-misconceptions}

The surrogate modeling approaches discussed above highlight deep learning's promise for multiscale problems. When trained with appropriate inductive biases on high-fidelity simulations, these models serve as faithful interpolation techniques bridging different resolutions and physical regimes. However, a common misconception emerges when similar techniques are applied to observational data with claims of achieving superresolution or enhanced signal-to-noise ratio (SNR) \citep[e.g.,][]{Vojtekova2021}. Although these models can be useful for cursory inspection and fitting \citep{Pal2024}, one should be cautious about claims of improved information content.

Deep learning models cannot create information that does not exist in the input data. When a network denoises a spectrum or enhances its resolution, it is not improving statistical constraining power---it is projecting learned patterns from training data onto observations. Consider a model trained to convert low-SNR to high-SNR spectra: Although the output may appear cleaner and yield tighter posterior constraints when fitted, these apparently improved constraints merely reflect the prior beliefs encoded in the training data. The danger lies in the opacity of this process---the implicit priors imposed by the neural network are difficult to trace or quantify, potentially leading to overconfident or biased scientific conclusions.

\subsection{Simulation-Based Inference}
\label{subsection:simulation-based-inference}

The multiscale modeling techniques discussed above enhance our ability to simulate astrophysical systems. These forward models now capture phenomena from cosmological structure formation to stellar atmospheres. However, modeling sophistication exacerbates a fundamental challenge: How do we perform statistical inference when comparing complex simulations with observational data?

Bayesian inference provides the framework for SBI. Given observations $\mathbf{d}$ and model parameters $\boldsymbol{\theta}$, we seek the posterior distribution $p(\boldsymbol{\theta}|\mathbf{d}) = p(\mathbf{d}|\boldsymbol{\theta}) p(\boldsymbol{\theta})/p(\mathbf{d})$, where $p(\mathbf{d}|\boldsymbol{\theta})$ is the likelihood, $p(\boldsymbol{\theta})$ is the prior, and $p(\mathbf{d})$ is the evidence. Although conceptually straightforward, evaluating this expression becomes intractable for astrophysical problems.

Astronomers traditionally compress data into summary statistics---two-point functions, power spectra, or bispectra---reducing data to manageable dimensions for analytical likelihood approximation \citep{Peebles1980}. However, even well-motivated statistics exhibit complex, non-Gaussian likelihood surfaces due to nonlinear physics, systematic effects, or finite samples. High-order statistics like the bispectrum captures non-Gaussian information but its likelihood deviates from Gaussianity in the nonlinear regime \citep{Martin2012}. The challenge intensifies for field-level inference---using full pixel information rather than compressed statistics.

The likelihood $p(\mathbf{d}|\boldsymbol{\theta})$ for an entire observed structure (e.g., the 3D galaxy distribution, 21-cm reionization map) is highly non-Gaussian.

\begin{marginnote}[]
\entry{Field-level inference}{Bayesian inference on full stochastic field rather than summary statistics; exploits complete information but requires neural density estimation}
\end{marginnote}

SBI leverages our ability to run forward simulations. Rather than deriving analytical likelihoods, SBI methods learn density estimators with neural networks that approximate intractable likelihoods or posteriors directly from simulated data (see \textbf{Supplemental Text, Section D} for technical details on normalizing flows, diffusion models, and the distinction between neural likelihood and posterior estimation). These approaches handle both non-Gaussian likelihoods of summary statistics \citep{Jeffrey2021} and field-level inference \citep{Zhao2022,Lemos2024} in which no tractable likelihood exists. By training neural networks on simulation-observation pairs, SBI extracts information from data complexity without requiring analytical expressions.

\subsubsection{Early approaches and their limitations: approximate Bayesian computation and generative adversarial networks}
\label{subsubsection:early-approaches-and-their-limitations-abc-and-gans}

Although SBI has become standard terminology, earlier astronomical literature often referred to these methods as likelihood-free inference (LFI). This terminology is misleading---any Bayesian inference relies on the likelihood, as evident from Bayes' theorem. What LFI actually means is avoiding explicit analytical likelihood functions (like assuming Gaussianity). Instead, LFI methods work with implicit likelihoods encoded in forward simulators. Recognizing this distinction, the field has adopted SBI as the more accurate descriptor.

The pursuit of inference without explicit likelihoods predates modern deep learning. Approximate Bayesian computation (ABC), developed in the 1990s \citep{Tavare1997,Beaumont2002}, represents an early approach. The ABC algorithm is straightforward: Simulate many datasets using different parameter values drawn from the prior, then keep only those parameters that produce simulated data close to observations:
\begin{equation}
p_{\text{ABC}}(\boldsymbol{\theta}|\mathbf{d}) \propto p(\boldsymbol{\theta}) \cdot \mathbb{I}[\rho(\mathbf{d}_{\text{sim}}, \mathbf{d}) < \epsilon],
\end{equation}
where $\mathbf{d}_{\text{sim}}$ is data simulated with parameters $\boldsymbol{\theta}$, $\mathbf{d}$ is the observed data, $\rho$ measures distance between them, $\mathbb{I}[\cdot]$ is an indicator function (1 when true, 0 otherwise), and $\epsilon$ is the tolerance threshold. As $\epsilon \to 0$, this approximation converges to the true posterior.

ABC has been applied across astronomy, including constraining galaxy merger rates \citep{Cameron2012} and planet occurrence rates \citep{Hsu2020}. However, in high-dimensional spaces, randomly sampled simulations almost never land close to observed data \citep{Blum2009}. This forces uncomfortable choices: Either run millions of simulations (computationally prohibitive) or increase tolerance $\epsilon$ so much that the posterior becomes unreliable.

Deep learning offered hope for overcoming these limitations. Generative adversarial networks \citep[GANs;][]{Goodfellow2014} seemed promising for astronomical inference \citep{Mustafa2019}. To understand GANs, imagine two neural networks in competition: a forger (generator) that creates fake astronomical data, and a critic (discriminator) that tries to spot fakes. Through repeated rounds of this game, the forger learns to create increasingly realistic data. For cosmological inference, the idea was to train the forger to create data conditional on cosmological parameters, i.e., generate mock galaxy distributions.

\begin{marginnote}[]
\entry{Mode collapse}{generative model failure producing only distribution subsets and missing modes or generating overly narrow distributions; leads to incomplete posteriors}
\end{marginnote}

Unfortunately, GANs have a critical flaw for scientific applications: mode collapse \citep{Arjovsky2017}. The adversarial game encourages the forger to find shortcuts---once it discovers one way to fool the critic, it has little incentive to explore all possibilities. This is particularly problematic for astronomy in which multiple physical scenarios often explain the same observations. If the true posterior is multimodal---with distinct peaks corresponding to different parameter combinations that equally explain the data---GANs might capture only a few modes while missing others. Furthermore, studies applying GANs found they systematically underestimated uncertainties \citep{Rodriguez2018}. For a field in which properly quantifying uncertainty is crucial, this unreliability presents a significant barrier to adoption. These failures motivated the search for neural network approaches that could leverage deep learning's power while faithfully representing posterior uncertainties.

\subsubsection{Normalizing flows}
\label{subsubsection:normalizing-flows}

Advances enabling practical SBI came from neural density estimation methods that avoided adversarial training altogether. Normalizing flows \citep{JiminezRezende2015,Durkan2019} provided stable, mode-preserving density estimation with exact likelihood computation. Unlike GANs, which only generate samples without modeling probability densities and suffer from mode collapse, normalizing flows offer both reliability and exact likelihood evaluation.

The core idea is to transform complex distributions into simple ones we understand---typically standard Gaussians---via neural networks. This transformation works both ways: complex to simple (to evaluate probabilities) or simple to complex (to generate samples). Consider an astronomical analogy: Large-scale structure evolved from nearly Gaussian initial fluctuations through gravitational instability. If we could run this evolution backward, we would transform today's non-Gaussian galaxy distribution back to its Gaussian origins. Normalizing flows learn similar bidirectional transformations from data. In fact, neural ODEs discussed earlier represent this principle---by learning continuous evolution from Gaussian to non-Gaussian distributions, they naturally provide the invertible transformations that define normalizing flows \citep{Chen2018}.

Mathematically, this uses the change of variables formula from calculus. When transforming a probability distribution, densities change according to
\begin{equation}
p_X(\mathbf{x}) = p_Z(f_\psi^{-1}(\mathbf{x}))|\det J_{f_\psi^{-1}}(\mathbf{x})|,
\end{equation}
where $p_X(\mathbf{x})$ is the complex distribution we want to model, $p_Z(\mathbf{z})$ is a standard Gaussian base distribution, $f_\psi$ is an invertible neural network (INN) with parameters $\psi$, and the Jacobian determinant $|\det J_{f_\psi^{-1}}|$ accounts for volume changes under the transformation. Because the network $f_\psi$ is invertible with efficiently computable Jacobians---constraints that make normalizing flows a special class of neural networks---hence, these are sometimes called INNs in astronomical literature \citep{Ksoll2020} (see \textbf{Supplemental Text, Section D} for details).

Training normalizing flows leverages our simulation capabilities. We generate parameter-data pairs by sampling parameters from our prior and running forward simulations. The network then learns to model the joint distribution $p(\boldsymbol{\theta}, \mathbf{d})$ or conditional distributions. This resembles fitting a Gaussian mixture model \citep{Dempster1977} but with far more flexibility---though Gaussian mixtures combine fixed Gaussian components, neural networks learn arbitrary transformations to capture complex distribution shapes. Neural networks can learn either the likelihood $p(\mathbf{d}|\boldsymbol{\theta})$ (neural likelihood estimation) or the posterior $p(\boldsymbol{\theta}|\mathbf{d})$ directly (neural posterior estimation), offering complementary approaches for inference (see \textbf{Supplemental Text, Section D}).

\subsubsection{Diffusion models}
\label{subsubsection:diffusion-models}

The requirement for invertibility and tractable Jacobians constrains normalizing flow architectures. Each layer must maintain invertibility while computing Jacobians efficiently, limiting expressiveness and requiring many layers to build powerful models. Diffusion models emerged as a solution \citep{Sohl-Dickstein2015,Ho2020}, relaxing these architectural constraints while proving effective for high-dimensional data like images or 3D density fields. Rather than normalizing flows' direct transformations, diffusion models take a gradual approach inspired by physical diffusion processes. Like ink diffusing in water, they reverse the process of adding noise to data. By learning to reverse this corruption, we can generate new samples from pure noise (see \textbf{Supplemental Text, Section D}).

Mathematically, the forward process adds calibrated Gaussian noise over $T$ time steps,
\begin{equation}
q(\mathbf{x}_t|\mathbf{x}_{t-1}) = \mathcal{N}(\mathbf{x}_t; \sqrt{1-\beta_t}\mathbf{x}_{t-1}, \beta_t\mathbf{I}),
\end{equation}
where $\mathbf{x}_t$ is data at time step $t$, $\beta_t$ controls noise level, $\mathcal{N}$ denotes a Gaussian distribution, and $\mathbf{I}$ is the identity matrix. The signal attenuates by $\sqrt{1-\beta_t}$ while Gaussian noise with covariance $\beta_t\mathbf{I}$ is added. After sufficient steps (typically hundreds), original data converge to standard Gaussian: $\mathbf{x}_T \sim \mathcal{N}(0, \mathbf{I})$. The neural network learns to reverse this by predicting and removing noise at each step---given noisy data, it estimates the less noisy previous version. Iterating from pure noise generates new samples following the training distribution, enabling synthesis of complex astronomical data like galaxy images \citep{Smith2022}, ISM tomography \citep{Xu2023}, and reionization fields \citep{Zhao2023}.

This denoising approach offers key advantages over GANs. Each training step involves straightforward denoising rather than unstable adversarial games, ensuring better coverage of target distributions. Studies show diffusion models more faithfully capture complex astrophysical fields with less mode collapse than GANs \citep{Zhao2023}.

Although standard diffusion models lack direct likelihood computation owing to their stochastic nature, \citet{Song2020} showed that for diffusion processes described by stochastic differential equations there exists a corresponding deterministic ODE---the probability flow ODE---that produces the exact same probability distributions over time. By learning the score function $\nabla_{\mathbf{x}} \log p(\mathbf{x})$, which represents the gradient pointing toward regions of higher probability, one can construct this deterministic ODE. Because it is deterministic rather than stochastic, standard change of variables formulas apply, enabling exact likelihood computation just like normalizing flows. This revelation has opened doors for SBI applications with diffusion models, though their adoption in astronomy remains limited compared to normalizing flows, partly due to their computational cost: The sequential steps make them slower than single-pass normalizing flows.

Currently, the different neural density estimations thus occupy complementary niches in astronomical applications. Normalizing flows provide fast and exact likelihoods through explicit change of variables, making them ideal for applications requiring time-critical inference. Diffusion models offer superior architectural flexibility and training stability, excelling at modeling complex multimodal distributions and generating high-quality synthetic data \citep{Xu2023}, despite their sampling speed limitations.

\subsubsection{Flow matching}
\label{subsubsection:flow-matching}

Most recently, flow matching has emerged as a unifying framework that bridges normalizing flows and diffusion models \citep{Albergo2022,Lipman2022,Liu2022}. Building on the continuous-time formulation of neural ODEs but sidestepping their computational challenges, flow matching directly learns the velocity field that transports probability mass from noise to data. Flow matching uses simple regression: Given pairs of noise and data points, a neural network learns to predict the velocity connecting them, effectively learning how points should move to transform a Gaussian distribution into the data distribution.

This approach inherits the theoretical foundations of neural ODEs---including exact likelihood computation through continuous change of variables---while being easier to train. Once learned, the velocity field defines a deterministic ODE that can be integrated for sampling, combining the architectural flexibility of diffusion models with the probabilistic rigor of normalizing flows. Though still underexplored in astronomy, flow matching's balance of expressiveness, computational efficiency, and theoretical guarantees positions it as a path forward that may supersede many current approaches. \textbf{Table~\ref{table3}} summarizes the complementary strengths and weaknesses of various neural density estimation approaches.

\begin{table}[h]
\tabcolsep3pt
\caption{Comparison of neural density estimation methods for astronomical applications}
\label{table3}
\begin{center}
\begin{tabular}{@{}l|c|c|c|c@{}}
\hline
\textbf{Property} &\textbf{GANs} &\textbf{Normalizing Flows} &\textbf{Diffusion Models} &\textbf{Flow Matching}\\
\hline
Likelihood evaluation &\ding{55} &\checkmark &\ding{55}/\checkmark$^{\rm a}$ &\checkmark\\[2pt]
Architectural flexibility &\checkmark &\ding{55} &\checkmark &\checkmark\\[2pt]
Training stability &\ding{55} &\checkmark &\checkmark &\checkmark\checkmark\\[2pt]
Sample quality &\checkmark &\checkmark &\checkmark\checkmark &\checkmark\checkmark\\[2pt]
Sampling speed &\checkmark\checkmark &\checkmark\checkmark &\ding{55} &\checkmark\\
\hline
\end{tabular}
\end{center}
\begin{tabnote}
$^{\rm a}$Only via subset of score-based models.\\
Abbreviation: GANs, generative adversarial networks.
\end{tabnote}
\end{table}

\subsubsection{Illustrative examples: from exoplanets to cosmological fields}
\label{subsubsection:illustrative-example-exoplanet-atmospheres-gravitational-waves-and-cosmological-fields}

These neural density estimation capabilities prove useful across diverse astronomical domains. For exoplanet atmospheric inference, \citet{Vasist2023} demonstrated normalizing flows' practical advantages. Traditional Markov chain Monte Carlo (MCMC) requires repeatedly evaluating expensive radiative transfer models to constrain temperature profiles, chemical abundances, and cloud properties from emission spectra. Instead, they trained normalizing flows to learn $p(\boldsymbol{\theta}|\mathbf{d})$ directly from simulated spectra, providing full posterior distributions for new observations in seconds with comparable accuracy to nested sampling.

In gravitational wave astronomy, in which alerts require rapid electromagnetic follow-up, \citet{Green2020b} showed normalizing flows enable real-time parameter estimation that would be impossible with traditional methods. Similarly, \citet{Zhang2021} demonstrated that amortized neural posterior estimation can infer binary microlensing parameters in seconds rather than the hours or days required by traditional grid searches and MCMC, which is critical for the thousands of events expected from the Nancy Grace Roman Space Telescope.

SBI has also catalyzed a paradigm shift in cosmological inference. Traditional analyses compress observations to summary statistics---power spectra, correlation functions, peak counts---inevitably discarding information. Even sophisticated statistics lack tractable likelihood expressions when capturing non-Gaussian features \citep{Jeffrey2021}. SBI transcends these limitations by enabling field-level inference that exploits full spatial information. For 21-cm reionization, \citet{Zhao2022} showed that 3D CNN approaches with SBI outperform conventional power spectrum MCMC in both accuracy and computational efficiency. Similarly, \citet{Lemos2024} applied field-level SBI to galaxy clustering, extracting non-Gaussian information inaccessible to traditional analyses.

These field-level approaches leverage different neural architectures depending on data structure. CNNs process gridded density fields, whereas GNNs naturally handle discrete galaxy catalogs and stellar streams for which order is arbitrary \citep{Lee2024}. Whether using CNNs for fields or GNNs for point clouds, the extracted features or latent representation $\mathbf{d}$ are then fed into normalizing flows that learn either $p(\boldsymbol{\theta}|\mathbf{d})$ for direct posterior estimation or $p(\mathbf{d}|\boldsymbol{\theta})$ for likelihood estimation, enabling efficient Bayesian inference.

\subsubsection{A cautionary note: on the black box critique}
\label{subsubsection:on-the-black-box-critique}

Despite these successes, SBI methods face legitimate scrutiny. High-dimensional distribution approximation with neural networks remains approximate, with no guarantee of posterior accuracy. The rapid adoption and subsequent abandonment of GANs in astronomy serves as a cautionary tale. Although most SBI studies validate against established methods like MCMC \citep{Green2020b,Vasist2023}, these are typically spot checks rather than comprehensive validation. When posterior accuracy is critical---as in cosmological parameter inference---rigorous blind tests across diverse systematic effects are essential.

The common black box criticism, however, deserves more nuanced consideration. Critics argue that neural networks obscure the relationship between data and parameters, but this concern applies equally to all modern inference pipelines that rely on complex simulations with unknown systematics. Furthermore, when SBI learns the likelihood of summary statistics while capturing their non-Gaussian features, it should produce more robust posteriors than methods that incorrectly assume Gaussian likelihoods.

The real challenge is not interpretability but testability. Even established summary statistics often defy simple intuition, yet we trust them because we can systematically probe how different physical effects manifest in their measurements \citep{Foreman2020}. Interpretability itself is ill-defined---what seems intuitive to one researcher may be opaque to another. The question is not whether our statistics are intuitively understandable, but whether we can validate their behavior under different conditions.

For field-level inference specifically, the trade-offs remain unclear. Although utilizing full spatial information could improve constraints, flexible neural networks risk learning simulation-specific features. For example, \citet{Villaescusa-Navarro2022} demonstrated that models trained on one simulation that successfully inferred cosmological parameters failed when applied to other simulations, revealing they had learned simulation artifacts rather than universal physics. In weakly non-Gaussian regimes, well-chosen summary statistics may already capture most cosmological information \citep{Cheng2020}. However, for inherently complex problems---such as exoplanet atmospheric retrieval or Milky Way dynamical modeling---SBI's ability to amortize expensive simulations may justify the additional validation burden. Ultimately, whether the potential gains justify the additional complexity depends on the specific application and the reliability of underlying simulations.

\subsection{Anomaly and Outlier Detection}
\label{subsection:anomaly-and-outlier-detection}

The neural density estimators that enable SBI offer an additional capability: systematic identification of outliers through probabilistic assessment. By learning the distribution $p(\mathbf{d})$ of normal astronomical objects, these methods quantify how typical or atypical any given observation appears---identifying anomalies as observations in which $-\log p(\mathbf{d}) > \tau$ for some threshold $\tau$ chosen to control the false positive rate. The same neural density estimators discussed for SBI---normalizing flows, diffusion models, and flow matching---naturally provide this capability through their likelihood computation. Objects with sufficiently low probability are flagged as outliers, offering a principled probabilistic measure rather than arbitrary distance metrics (see \textbf{Supplemental Text, Section D}). This approach requires only normal training data without labeled anomalies, making it particularly valuable for discovering unknown classes of astronomical phenomena where advancements often emerge from identifying objects that defy existing classification schemes.

\subsubsection{Variational autoencoders}
\label{subsubsection:variational-autoencoders-vaes}

Although generative models can learn distributions directly in observation space, the curse of dimensionality makes this computationally expensive. Astronomical objects, however, possess low intrinsic dimensionality---a galaxy's observable properties emerge from perhaps dozens of underlying parameters rather than tens of thousands of pixels. This motivates learning compact representations before density estimation. The SBI approaches discussed earlier already leverage this principle---CNNs extract hierarchical features, GNNs learn graph embeddings, and Transformers compute attention-based representations, compressing observations before feeding them to density estimators.

Variational autoencoders (VAEs) offer an alternative with a distinct philosophy: They explicitly optimize for representations that are both informative and probabilistically structured \citep{Kingma2013}. Unlike standard autoencoders that simply compress and reconstruct data, VAEs model the encoding process probabilistically:
\begin{equation}
q_\phi(\mathbf{z}|\mathbf{x}) = \mathcal{N}(\boldsymbol{\mu}_\phi(\mathbf{x}), \boldsymbol{\sigma}^2_\phi(\mathbf{x})),
\end{equation}
where the encoder network with parameters $\phi$ outputs both mean $\boldsymbol{\mu}_\phi(\mathbf{x})$ and variance $\boldsymbol{\sigma}^2_\phi(\mathbf{x})$ for each latent dimension $\mathbf{z}$. This means each input maps not to a single point but rather to a distribution in latent space. Training balances reconstruction accuracy while keeping these latent distributions close to a standard Gaussian prior $p(\mathbf{z}) = \mathcal{N}(0, \mathbf{I})$. This dual constraint creates organized representations in which similar objects cluster near the origin. For anomaly detection, this provides two signals: both objects with extreme latent positions (far from the Gaussian origin) and poor reconstruction despite typical latent values indicate outliers---capturing different types of astronomical anomalies.

\subsubsection{Illustrative examples: from spectroscopic surveys to time-domain astronomy}
\label{subsubsection:illustrative-examples-from-spectroscopic-surveys-to-time-domain-astronomy}

Anomaly detection has become increasingly vital across astronomy, from identifying peculiar galaxies and quasars to discovering new classes of transients and uncovering systematic errors in large surveys. We highlight two representative examples that demonstrate the power of neural density estimation and VAE approaches.

\citet{Liang2023} applied the autoencoder-normalizing flow combination to $\sim$500,000 optical spectra from the Sloan Digital Sky Survey. Their two-stage approach compressed high-dimensional spectral data into compact representations, then modeled their distribution with normalizing flows. This discovered diverse anomalies spanning interesting objects (Wolf--Rayet stars, active supernovae, galaxy mergers) and data quality issues (blended objects, misclassified stars), illustrating how anomaly detection serves both discovery and quality control.

For time-domain astronomy, where the Rubin Observatory will discover millions of transients annually \citep{Ivezic2019}, real-time anomaly detection enables targeted follow-up of rare events. \citet{Villar2021} applied VAEs with RNN backbones to simulated Rubin Observatory light curves, identifying superluminous supernovae, pair-instability supernovae, and intermediate luminosity transients before peak brightness. The probabilistic anomaly scores from VAEs provide principled rankings for spectroscopic follow-up in an era of overwhelming time series data with the Rubin Observatory.

\subsection{Foundation Models}
\label{subsection:foundation-models}
    
The techniques discussed thus far---multiscale modeling, SBI, and anomaly detection---represent natural extensions of classical machine learning paradigms. Although these applications demonstrate deep learning's power, they remain conceptually grounded in traditional statistical approaches: learning mappings from training data to make predictions on similar test data.

Recent advances in machine learning have revealed that scaling neural networks to billions of parameters leads to emergent capabilities---models develop general representations that transfer across diverse tasks without task-specific training \citep{Brown2020}. This transferability has led to the term foundation models: large-scale models trained on broad data that serve as foundations for specialized applications through fine-tuning. Unlike smaller neural networks that suffer from catastrophic forgetting---losing previously learned capabilities when trained on new tasks \citep{McCloskey1989}---foundation models maintain broad capabilities while adapting to new domains.

\begin{marginnote}[]
\entry{Catastrophic forgetting}{neural networks abruptly losing previously learned capabilities when trained on new tasks, overwriting old knowledge with new training data}
\end{marginnote}

The hallmark of foundation models is zero-shot and few-shot learning. Zero-shot learning means performing entirely new tasks without any task-specific training, whereas few-shot learning requires only a handful of examples. For instance, language models trained on general text can write poetry or answer scientific questions---tasks that the models were never explicitly trained for---by leveraging their broad understanding of language structure \citep{Brown2020}. For astronomy, this suggests models trained on diverse data might generalize to new phenomena by exploiting universal physical laws. Foundation models achieve strong performance with minimal task-specific data through self-supervised learning on diverse unlabeled data, enabling rapid adaptation to new tasks, which contrasts with traditional approaches requiring extensive labeled datasets for each application (see \textbf{Supplemental Text, Section E}).

\subsubsection{Self-supervised and cross-modal training strategies}
\label{subsubsection:self-supervised-and-cross-modal-training-strategies}

Foundation models learn transferable representations through self-supervised and cross-modal training strategies that extract meaningful patterns from unlabeled data---techniques that have proven useful in natural language processing. These self-supervised approaches embed domain knowledge through pretext tasks---auxiliary objectives that force meaningful learning without labels.

\begin{marginnote}[]
\entry{Self-supervised learning}{training without external labels using data's inherent structure; predicting masked regions, aligning modalities, or contrasting examples enables learning physical relationships from unlabeled data}
\end{marginnote}

Self-supervision creates learning signals from data structure itself. Masked autoencoding \citep{He2021} represents this approach. The objective becomes
\vspace{-0.1cm}
\begin{equation}
\mathcal{L}_{\text{MAE}} = \frac{1}{|\mathcal{M}|}\sum_{i \in \mathcal{M}} ||\mathbf{x}_i - f_\theta(\mathbf{x}_{\setminus \mathcal{M}})||^2, \vspace{-0.2cm}
\end{equation}
where $\mathcal{M}$ denotes masked indices, $\mathbf{x}_{\setminus \mathcal{M}}$ represents unmasked portions, and $f_\theta$ reconstructs masked regions from context. Although seemingly arbitrary compared to physics-based constraints, this task encodes deep astronomical knowledge. For spectra, for example, predicting masked wavelength regions requires understanding atomic physics: If iron lines at 4000 \AA{} are strong, the network must learn that iron lines at 5000 \AA{} should appear with correlated strengths. The model discovers which atomic species contribute to different spectral regions and how their relative intensities encode temperature, density, and composition.

Beyond single-modality learning, cross-modal training addresses the limitation that learning from one observational technique risks capturing instrumental artifacts rather than physical properties. Cross-modal learning requires different modalities to encode the same information into shared representations, yielding more robust features. Consider a galaxy observed through photometry and spectroscopy: Both broadband colors and detailed spectral features encode stellar populations, metallicities, and star-formation histories. Cross-modal architectures discover unified descriptions by training separate encoders to map these observations into a common latent space. This constraint forces networks to discover features capturing underlying physics rather than modality-specific artifacts.

Contrastive learning provides another effective training framework. Instead of reconstruction, it learns by pulling together different views of the same object while pushing apart different objects \citep{Hadsell2006}. The encoders produce latent representations $\mathbf{z}$ for each modality. Given an image-spectrum pair from the same galaxy (positive) and pairs from different galaxies (negative), the loss becomes
\vspace{-0.1cm}
\begin{equation}
\mathcal{L}_{\text{contrast}} = -\log \frac{\exp(\text{sim}(\mathbf{z}_{\text{image}}, \mathbf{z}_{\text{spectrum}})/\tau)}{\sum_j \exp(\text{sim}(\mathbf{z}_{\text{image}}, \mathbf{z}_j)/\tau)},
\end{equation}
where $\mathbf{z}_{\text{image}}$ and $\mathbf{z}_{\text{spectrum}}$ are latent representations from respective encoders, $\text{sim}(\cdot,\cdot)$ measures similarity, and $\tau$ controls discrimination strength. The numerator increases when the representations align, whereas the denominator normalizes over negative examples.

\begin{marginnote}[]
\entry{Contrastive learning}{training by pulling similar examples together while pushing dissimilar ones apart; learns which observations should be close in latent space}
\end{marginnote}

\subsubsection{Illustrative examples: bridging synthetic and observed spectra}
\label{subsubsection:illustrative-examples-bridging-synthetic-and-observed-spectra}

\citet{OBriain2021} developed a model that treats synthetic models and observations as distinct modalities requiring alignment. The network learns bidirectional mappings while enforcing shared latent representations, discovering systematic corrections that calibrate theoretical predictions to observational reality. By training on both domains, the method identifies and corrects for theoretical inadequacies, instrumental signatures, and continuum normalization differences that have historically limited model-based spectral analysis. Similarly, \citet{Zhao2025} applied cross-domain masked autoencoding to stellar spectra from Gaia XP and LAMOST, demonstrating that self-supervised pretraining improves stellar parameter estimation even with limited labels. Their model learns spectral physics by predicting masked wavelength regions, discovering atomic line correlations without explicit supervision.

\citet{Parker2024} extended these ideas to cross-modal learning between images and spectra \citep[see also][]{Rizhko2025}. By training encoders to align galaxy images with corresponding spectra in shared representation space. The cross-modal constraint forces identification of features common to both modalities; stellar populations manifest as both image colors and spectral absorption features.

\subsubsection{A cautionary note: the information trade-off in foundation models}
\label{subsubsection:a-cautionary-note-the-gap-between-promise-and-reality}

Despite the promise of foundation models, the limited improvements observed thus far raise questions about their practical gains in astronomy \citep{Zhao2025}. A common misconception claims that foundation model pretraining inherently improves downstream tasks. Self-supervised pretraining optimizes for reconstruction rather than task-specific performance. Like principal component analysis which minimizes global reconstruction error rather than preserving task-relevant information, self-supervised learning loses information by design \citep{Ruan2025} and cannot supersede supervised methods when ample labeled data are available. For data-rich tasks like photometric redshift estimation, simple supervised learning remains optimal.

This reality contrasts sharply with the terminological inflation in recent astronomical literature. The term foundation model originally denoted models capable of zero-shot or few-shot transfer across diverse tasks. Yet much astronomical work bearing this label merely applies Transformer architectures to specific tasks, conflating architectural choices with foundational capabilities. True foundation models require transferable knowledge and few-shot learning, not specific architectures.

Indeed, the capacity for generalization in overparameterized models is quite general (see Section~\ref{subsection:deep-learning-direction}), emerging from phase transitions in which models discover generalizable solutions instead of memorize \citep{Bahri2020,Power2022}. This occurs across architectures---from linear models \citep{Hastie2019} to Gaussian processes \citep{Canatar2021}---and not exclusively in Transformers. The existing astronomical cross-modal approaches discussed, though valuable for multisource learning, remain task-specific models lacking the versatility shown in modern day large language models (LLMs).

\begin{marginnote}[]
\entry{Neural scaling laws}{empirical power-law relationships between model performance and scale factors; performance improves predictably, enabling prediction of capabilities before training}
\end{marginnote}

Still, foundation models prove valuable when labeled data are scarce, which is precisely where traditional methods fail. In these regimes, emerging evidence suggests genuine progress. Neural scaling laws---predictable performance improvements with model and data size \citep{Kaplan2020}---appear across astronomical applications in images \citep{Walmsley2024}, spectra \citep{Rozanski2025b}, and time series \citep{Pan2024b}. Studies have shown that domain-specific pretraining in astronomical data provides benefits even when pretraining and target domains differ, indicating possible cross-domain knowledge transfer in the label-poor regime \citep{Walmsley2024,Rozanski2025a}.

\subsection{Reinforcement Learning and Workflow Optimization}
\label{subsection:reinforcement-learning-and-workflow-optimization}

Although the deep learning applications discussed thus far focus on understanding and modeling astronomical data, many astronomical challenges require optimizing sequences of decisions---from telescope scheduling to adaptive optics control. Reinforcement learning (RL) addresses this complementary challenge: learning optimal decision strategies through trial and error, where actions taken now affect future opportunities.

RL discovers strategies by interacting with environments and receiving feedback. A state encapsulates all relevant information about the current situation; for a telescope, this might include target positions, current pointing, weather conditions, and instrument status. The algorithm observes this state and chooses an action that changes the environment, leading to a new state. Consider DeepMind's AlphaGo, which learned Go by evaluating board positions (states) and their eventual outcomes \citep{Silver2016}. The key insight of RL is that neural networks can learn to evaluate complex states and select actions that maximize long-term success rather than immediate gain.

In RL, a policy $\pi$ maps states to actions:
\begin{equation}
p(a|s) = \pi_\theta(s),
\end{equation}
where $p(a|s)$ is the probability of taking action $a$ in state $s$, and $\theta$ represents the neural network parameters. The policy outputs probabilities over possible actions, enabling exploration of different strategies. This represents a philosophical shift from engineering explicit rules to learning implicit strategies. The network optimizes expected cumulative reward,
\begin{equation}
\pi^* = \arg\max_\pi \mathbb{E}\left[\sum_{t=0}^{\infty} \gamma^t r_t \bigg| \pi\right].
\end{equation}
The optimal policy $\pi^*$ maximizes total reward over trajectories generated by following policy $\pi$, where $r_t$ quantifies success at time $t$ and discount factor $\gamma < 1$ ensures convergence by exponentially downweighting future rewards, prioritizing immediate over distant outcomes. Crucially, RL requires explicit, quantifiable rewards---the system must know what constitutes success.

\subsubsection{Illustrative examples: telescope scheduling}
\label{subsubsection:illustrative-examples-telescope-scheduling-adaptive-optics-and-gravitational-wave-interferometers}

These RL principles find natural applications in astronomy's operational challenges. Telescope scheduling illustrates the complexity: hundreds of interacting variables including weather patterns, instrument constraints, and competing science goals. At its core, this resembles the traveling salesman problem: finding optimal sequences to visit celestial targets while minimizing slew time and maximizing efficiency. However, astronomical scheduling adds layers of complexity. Visibility windows, changing atmospheric conditions, instrument reconfigurations, and time-dependent priorities transform this into a dynamic challenge. Traditional approaches rely on priority queues and heuristic rules. Although sensible, such rules cannot capture full system complexity. Deep RL replaces these handcrafted rules with learned policies that process complete system states.

\citet{Cao2025} demonstrated this approach by representing scheduling as a graph in which targets are nodes and transition times are edges. The neural network processes both node information (target properties) and edge information (transition costs) to learn optimal sequences. The reward function balances SNRs with observation completeness. Through thousands of simulated episodes, the network discovers patterns human operators might miss, successfully scheduling real observations when deployed at the Xinjiang Astronomical Observatory's Muztagh Observation Station. Similarly, \citet{Zhang2025} applied RL to distributed telescope arrays, achieving improved performance over traditional heuristics when handling dynamic constraints across geographically separated sites.

\subsubsection{Illustrative examples: adaptive optics and gravitational wave interferometers}
\label{subsubsection:illustrative-examples-adaptive-optics-and-gravitational-wave-interferometers}

Adaptive optics presents another application. Classical systems react to current wavefront errors using linear models, which is adequate for static corrections but insufficient when accounting for dynamic atmospheric turbulence. Deep RL policies instead process temporal sequences of wavefront measurements. The reward is explicit: Strehl ratio or other image quality metrics directly quantify correction performance. By analyzing patterns across measurements, networks anticipate turbulence evolution, enabling predictive rather than reactive correction.

\citet{Nousiainen2022} employed RL for adaptive optics control, using CNNs to extract spatial representations from deformable mirror and wavefront sensor data \citep[see also][]{Landman2021}. The training process collects real-time episodes by running the policy in actual control loops, enabling the network to discover optimal strategies through system interaction. The approach achieved notable improvement in coronagraphic contrast compared to traditional integrator control and was validated through simulations for 8-m and 40-m telescopes and laboratory experiments at Steward Observatory.

RL has also been applied to control gravitational wave detectors themselves. \citet{Buchli2025} used RL to control the Laser Interferometer Gravitational-Wave Observatory (LIGO) mirror stabilization system, where quantum-limited sensitivity requires precise control against seismic disturbances. Traditional linear controllers struggle with complex, frequency-dependent noise coupling in the 10--30-Hz band. The neural network observes sensor signals, adjusts actuator commands, and receives rewards for minimizing noise in target frequency bands, extending the LIGO's astrophysical reach.

\vspace{-0.5cm}
\subsection{Large Language Models and Agentic Research}
\label{subsection:large-language-models-and-agentic-research}

Although RL excels at optimizing well-defined objectives---from telescope scheduling to detector control---many astronomical discoveries emerge from open-ended exploration that resists scalar rewards. When astronomers examine anomalous spectra or unexpected correlations, they engage in functional generation and creative reasoning guided by intuition and accumulated knowledge. This broader challenge motivates exploration of LLMs and their potential for autonomous scientific discovery.

The astronomy community has developed specialized models tailored to our domain. Recent efforts demonstrate that open-weight LLMs, when fine-tuned on astronomical literature and data, can achieve recall performance rivaling flagship proprietary models while reducing computational costs \citep{deHaan2025}. Although LLMs as scientific copilots raise profound questions about the identity of astronomers \citep{Ting2025b} and education \citep{Ting2025c}, their technical capabilities warrant serious consideration.

The concept of self-driving laboratories, now actively explored in materials science \citep{Szymanski2023} and chemistry \citep{Boiko2023}, suggests possibilities for astronomy. These systems mine synthesis procedures from tens of thousands of publications to guide autonomous experimentation. Beyond information retrieval, agentic research represents a more ambitious vision \citep{Wang2025b}: AI systems that act as autonomous agents, planning, executing, and adapting research strategies without human intervention. These agents can serve as walkers in functional space, navigating territories where no clear reward function exists and pursuing lines of inquiry not because they maximize any predefined metric but because they open new theoretical possibilities. Unlike RL with explicit scalar rewards, these agents employ reflection mechanisms to interpret complex evaluation metrics and identify why approaches succeed or fail, iteratively selecting promising directions, generating hypotheses through learned patterns, and evaluating candidates via simulations or analysis. This enables exploration of scientific territories without predefined optimization targets, progressively accumulating strategic knowledge about which approaches work for different observational signatures.

However, the Moravec paradox tempers this vision: Though AI excels at high-level reasoning tasks traditionally considered difficult (like proving mathematical theorems), it struggles with sensorimotor skills, common sense, and perception that humans find trivial. Although AI agents have achieved gold-medal-level performance at the International Olympiad on Astronomy and Astrophysics \citep{Pinheiro2025}, using AI as completely autonomous agents rather than copilots requires more holistic capabilities. Human-in-the-loop systems tolerate errors that fully autonomous systems cannot afford. Basic capabilities like accurately reading plots remain imperfect, falling behind human performance \citep{Wang2024}.

Most critically, astronomy faces a unique challenge that distinguishes it from fields where self-driving laboratories have shown promise: Our bottleneck is not functional generation but verification. Proposing new dark matter models or modified cosmological parameters is straightforward; testing them requires costly simulations, statistical inference, and detailed understanding of observational biases. This asymmetry fundamentally shapes how autonomous reasoning systems must operate in our field, suggesting that near-term applications will focus on building the ecosystem, that is, developing agent-ready tools and interfaces that allow LLM agents to interact meaningfully with astronomical data, simulations, and existing analysis pipelines rather than overpromising on fully autonomous discovery.

Although fully autonomous discovery may remain distant, LLM-based agents are already demonstrating value in well-scoped astronomical tasks that leverage their reasoning capabilities within controlled domains. Two recent examples illustrate complementary applications: navigating physical model spaces and discovering algorithmic solutions.

Mephisto \citep{Sun2025} employs LLM-based agents that use tree search to navigate different physical models for JWST (James Webb Space Telescope) spectral energy distribution fitting. Unlike traditional parameter optimization that tweaks values within fixed models, these agents reason about model-data discrepancies and propose structural changes, switching from dust attenuation to AGN models when spectral features suggest alternative physics. Through iterative refinement, the system accumulates strategic knowledge about which model modifications improve fit quality for different observational signatures. Applications demonstrate practical value: \citet{Wei2025} used Mephisto to analyze Zangetsu, an unusual quiescent ultradiffuse galaxy whose extreme properties made conventional fitting unreliable, whereas \citet{Wang2025a} applied it to validate luminous quasar results.

A complementary application addresses algorithm discovery rather than model selection. \citet{Wang2025b} deployed LLM agents for automated gravitational wave detection algorithm discovery. Through iterative cycles of generation, evaluation, and refinement guided by LLM heuristics, the system progressively discovers algorithms balancing computational efficiency with detection performance. This demonstrates how agents can autonomously navigate solution spaces beyond simple parameter tuning, accumulating knowledge about algorithmic strategies that emerge and propagate through the search process.

\section{FINAL THOUGHTS}
\label{section:final-thoughts}

This review necessarily captures a field in motion, frozen at a specific moment---summer 2025 when this article was written. What we present as cutting-edge today may be routine tomorrow; techniques we overlook may prove transformative. This temporal limitation is inherent to reviewing any fast-moving field but particularly acute for deep learning, where capabilities advance monthly rather than yearly. Furthermore, given tough page and number-of-reference constraints imposed by Annual Reviews, this review inevitably omits substantial literature. Rather than attempt comprehensive coverage, we aimed to balance topics across astronomical subfields while highlighting work with broad methodological reach---the basis functions from which future applications might be constructed. By surveying this diverse landscape, we can glimpse potential future impacts, however speculative they may be.

Despite these limitations, patterns emerge from examining where deep learning has gained traction versus where it remains peripheral. Understanding these patterns---and the underlying bottlenecks they reflect---helps identify domains in which future advances might prove most transformative.

\subsection{Where Deep Learning Might Have the Greatest Impact}
\label{subsection:where-deep-learning-might-have-the-largest-impact}

Astronomical knowledge gaps typically arise from four sources:

\begin{itemize}
\item \textbf{(a) Sample size limitations:} Data volume imposes constraints, both sampling noise and insufficient rare objects. Although instrumentation advances address this through astronomical surveys, this limitation is largely independent of deep learning.

\item \textbf{(b) Fundamental theory:} The gaps stem from incomplete theoretical frameworks. Tensions among high-precision cosmological measurements illustrate that even perfect analyses cannot resolve questions rooted in fundamental physics. LLM-based agents might eventually contribute, though this remains speculative.

\item \textbf{(c) Information extraction:} There are challenges involved in extracting information from high-dimensional, non-Gaussian data. The SBI and anomaly detection techniques reviewed above aim to address this gap.

\item \textbf{(d) Simulation systematics:} Nonlinear astrophysical systems require numerical simulations that suffer from subgrid approximations and synthetic-observation gaps. Multiscale modeling and foundation models offer pathways forward.
\end{itemize}

\begin{figure}[ht!]
\centering
\includegraphics[width=\textwidth]{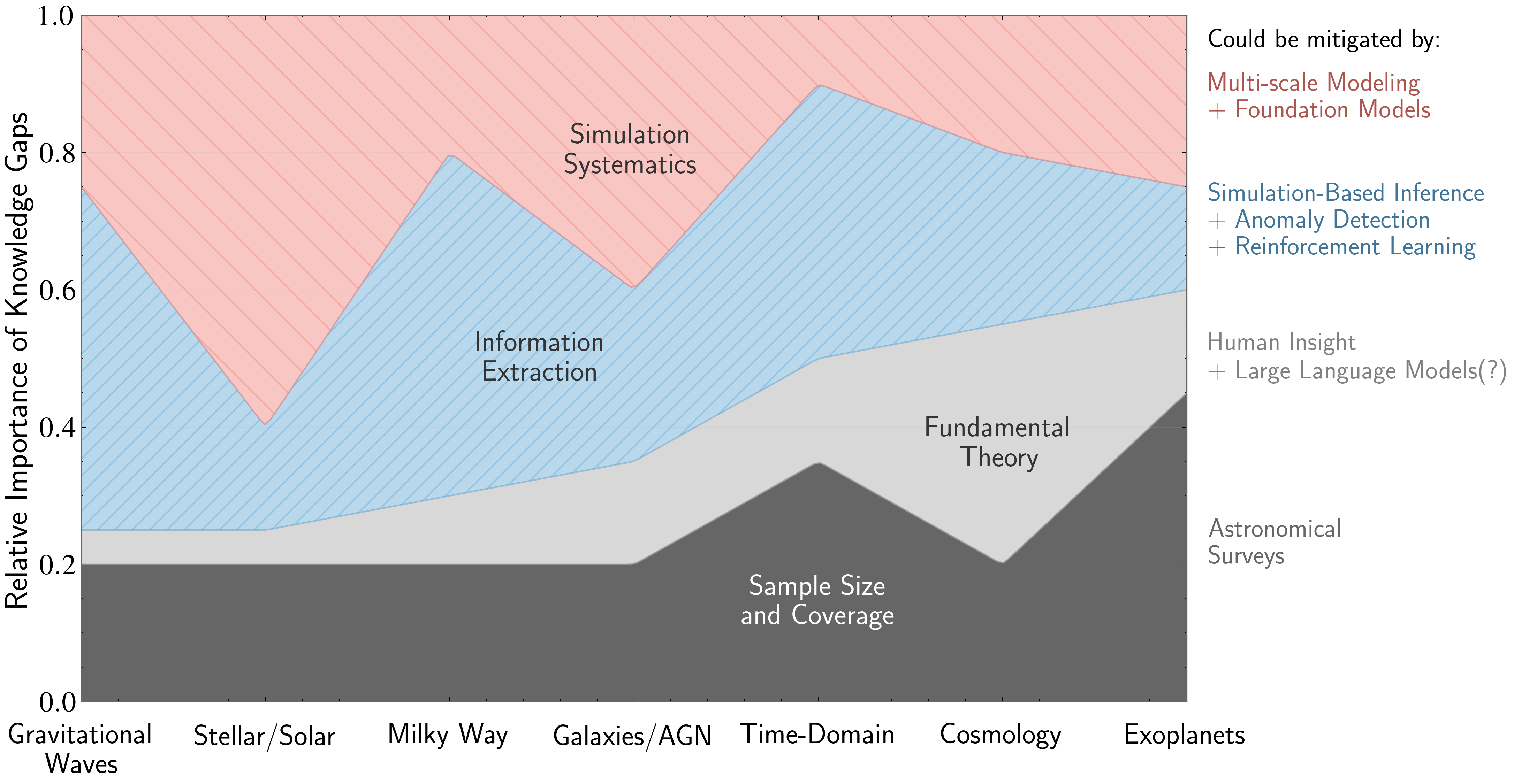}
\caption{Knowledge gaps across astronomical domains. Bands show the rough relative importance of four limitation types: sample size (dark gray), fundamental theory (light gray), information extraction (blue), and simulation systematics (red). Domains are ordered by deep learning addressability, with hatching patterns indicating gaps potentially mitigated by deep learning. Right annotations show approaches for each limitation type. These proportions are approximate guides rather than precise quantifications. Abbreviations: AGN, active galactic nucleus; LLM, large language model.}
\label{fig3}
\vspace{-0.5cm}
\end{figure}

Deep learning primarily addresses points c and d. Understanding where closing these gaps would enable discoveries reveals how relative importance varies across domains (\textbf{Figure~\ref{fig3}}):

\begin{itemize}
\item \textbf{Gravitational waves:} Theory is well established; detector sensitivity steadily improves sample sizes. Bottlenecks include information extraction and simulation accuracy. Template-free searches in non-Gaussian noise, numerical relativity surrogates for rapid inference, and uncertainty quantification all benefit from deep learning when care is taken to avoid selection biases. These capabilities could reveal new source classes.

\item \textbf{Stellar astrophysics and solar physics:} Theory is somewhat mature; surveys provide ample data. The primary challenge is simulation accuracy. 3D non-LTE spectral emulators and neural radiative transfer surrogates mitigate synthetic-observation gaps. Physics-informed approaches tie models to observations. Information extraction from asteroseismic and solar data further benefits from deep learning. Operationally, flare and coronal mass ejection forecasting demonstrates the practical value of deep learning.

\item \textbf{Milky Way:} Theory is well established for equilibrium dynamics, though bar/spiral dynamics and ISM physics remain incomplete. Astrometric and spectroscopic surveys provide unprecedented volumes \citep{Abdurrouf2022} for deep learning applications. Information extraction likely dominates the utility of deep learning---stream discovery, chemical evolution, dust mapping, and potential inference through SBI. Simulation systematics matter for chemical evolution models and complex selection functions, which deep learning can mitigate.

\item \textbf{Galaxy evolution and AGNs:} Theory remains incomplete for feedback and jet physics. Sample sizes from surveys like DESI call for deep learning methods. Multiscale simulation systematics often dominate. Deep learning aids information extraction (deblending, integral field unit analysis, spectral synthesis) and provides radiative-transfer and magnetohydrodynamics emulators spanning scales.

\item \textbf{Time-domain astronomy:} Theory suffices for discovery decisions, though the physics of fast radio bursts and tidal disruption events remain unclear. Sample-size limitations from rare transients and labeling bottlenecks dominate. Information extraction becomes critical; self-supervised pretraining, few-shot learning, and explainable triage address these challenges. Domain-shift-robust models work across instruments. RL optimizes follow-up scheduling among millions of candidates.

\item \textbf{Cosmology:} Developing theory for the nature of dark energy and dark matter is an ultimate goal, but near-term probes are systematics-limited. Sample sizes from surveys are improving. Immediate challenges: information extraction (photometric redshift, shear calibration, deblending, intrinsic alignments) and simulation systematics for baryonic effects. Deep learning enables field-level inference, non-Gaussian mapping, cosmic microwave background delensing, and fast multiscale emulators. Strong lensing embodies these advances: Lens finding and dark matter constraints all benefit from deep learning.

\item \textbf{Exoplanet science:} Theory spans well-understood dynamics to uncertain formation. High-quality spectra remain scarce; radial velocity follow-up is limited, thus limiting the applications of deep learning. However, information extraction challenges include detrending, point spread function suppression, radial velocity extraction with stellar activity modeling, and multimodal microlensing analysis can benefit from deep learning. Atmospheric retrieval emulators with deep learning further address simulation bottlenecks.
\end{itemize}

Deep learning will most impact domains in which information extraction and simulation systematics dominate: gravitational waves, time-domain astronomy, stellar astrophysics, and Milky Way science. Even in cosmology, where theory is the limiting factor, mitigating systematics can yield near-term progress. Theory-limited fields await breakthroughs that deep learning cannot directly provide, though it may accelerate discovery of anomalies that inspire new theories.

\subsection{Looking Forward}
\label{subsection:looking-forward}

Polarized discourse surrounding deep learning in astronomy---from enthusiastic adoption to skeptical dismissal---persists, often generating more heat than light. Yet, having traced deep learning's evolution across multiple dimensions, we can ground these discussions in concrete developments. This review demonstrates how deep learning achieves generalizability while maintaining expressivity and, through careful incorporation of astronomical inductive biases, becomes increasingly data efficient (Section~\ref{section:methodological-foundations}), with great potential across diverse astronomical applications (Section~\ref{section:cross-cutting-techniques}).

\begin{itemize}
\item \textbf{For AI-skeptic researchers:} Although neural network-based methods have existed for decades, the advances presented in this review represent genuine progress beyond glorified curve fitting. The prevalent black box critique---that neural networks sacrifice understanding for performance---deserves nuanced consideration. The term interpretability itself, like the term AI, lacks precise definition. Indeed, as shown in Section~\ref{section:methodological-foundations}, the ability to inject nuanced inductive biases demonstrates that these methods can incorporate astronomical knowledge rather than ignore it.

Deep learning incorporates modeling decisions through architectural choices, and physics-informed networks directly incorporate conservation laws. These architectural choices represent modeling decisions as fundamental as any parametric assumption. Ultimately, interpretability often reduces to the confidence that methods generalize beyond training domains and remain robust to systematics. Viewed through this lens, many techniques in this review demonstrate meaningful progress toward these goals, backed by increasingly deeper theoretical understanding of neural networks.

\item \textbf{For AI-curious researchers:} Although this review highlights advances in modern deep learning, these methods complement rather than replace classical approaches. Linear regression and Gaussian processes remain well-studied tools with fully understood Bayesian properties. For many astronomical problems---particularly those with limited data or requiring rigorous uncertainty quantification---these classical methods remain optimal. Crucially, though classical methods provide tractable posteriors over model parameters, neural networks struggle with this fundamental aspect of uncertainty quantification of the models.

\begin{marginnote}[]
\entry{Bayesian neural networks}{neural networks treating weights as probability distributions rather than fixed values, enabling uncertainty quantification but with computational and approximation challenges}
\end{marginnote}

Despite advances in Bayesian neural networks and ensemble methods, obtaining reliable posterior distributions over millions of network parameters remains computationally prohibitive and theoretically challenging. Physics assumes parsimony, and the same principle applies to data modeling. Start simple, add complexity only when justified by data and performance. Applying deep learning simply because it is available, without clear justification, risks obscuring understanding and wasting computational resources while potentially introducing unnecessary failure modes.

\item \textbf{For AI-enthusiast researchers:} The history of AI shows repeated cycles of elevated expectations followed by periods of recalibration. Learning from this pattern, we must approach deep learning with rigor, honestly reporting both capabilities and limitations. This balanced approach represents the only path to building lasting trust in these methods. Just as statistics has become integral to astronomical research without replacing scientific inquiry, deep learning represents a powerful addition to our toolkit rather than an end in itself. The possibility of more agentic systems---ones that could navigate functional spaces, propose novel theoretical frameworks, or reveal unexpected connections---represents genuine potential. Yet such capabilities remain highly speculative, and we serve the field best by acknowledging this uncertainty instead of succumbing to hype.

For a field that has always pushed the boundaries of human knowledge by looking outward, the development of deep learning represents another frontier, this time looking inward at the nature of intelligence and discovery itself. The advances documented in this review mark not an end but a beginning. As we stand at this threshold, we should celebrate concrete achievements while maintaining both healthy skepticism about hype and openness to possibilities we cannot yet imagine. The Universe has surprised us before; perhaps our deep learning tools will learn to surprise us too.
\end{itemize}

\begin{summary}[SUMMARY POINTS]
\begin{enumerate}

\item Classical machine learning faces trade-offs among scalability, expressivity, and data efficiency due to the curse of dimensionality. Deep learning sidesteps these limitations, with adoption in astronomy matching classical methods by 2025.

\item Specialized neural architectures encode domain-specific assumptions: convolutional neural networks leverage translation invariance for images, Transformers capture long-range dependencies in spectra, and graph networks handle irregular galaxy distributions, while equivariant networks preserve physical symmetries.

\item Physics-informed neural networks incorporate conservation laws and differential equations as constraints, enabling models that generalize beyond training data while respecting known physics, balancing flexibility with physical consistency.

\item The black box critique conflates interpretability with trustworthiness. Modern architectures achieve transparency by encoding physical knowledge through design choices that represent modeling decisions as important as parametric assumptions.

\item Astronomical simulations span vast scales that no single simulation can capture. Neural networks learn mappings between resolutions using encoder-decoder architectures and neural differential equations, though superresolution claims often mistake learned priors for genuine information gain.

\item Simulation-based inference (SBI) using neural density estimators (normalizing flows, diffusion models, flow matching) enables Bayesian posterior estimation when likelihoods are intractable, extracting information from complex distributions including field-level cosmological analysis.

\item Probabilistic anomaly detection through variational autoencoders and density estimation systematically identifies outliers in large surveys. As astronomical knowledge deepens and surveys grow, discovering the unexpected becomes increasingly critical for advancements.

\item Foundation models aim to learn transferable representations from diverse unlabeled data, though current astronomical applications often conflate Transformer architectures with foundational capabilities. When sufficient labeled data exist, supervised methods remain optimal; foundation models' value emerges in label-scarce regimes.

\item Reinforcement learning (RL) optimizes operational workflows from telescope scheduling to adaptive optics control and gravitational wave detector stabilization, learning decision policies that outperform hand-crafted heuristics in complex, dynamic environments.

\item Large language model (LLM)-based agents show potential for semiautonomous research tasks, though full autonomy remains limited by Moravec's paradox, i.e., excelling at high-level reasoning while struggling with basic perception.

\item Astronomy differs from industrial artificial intelligence (AI): We leverage strong physical priors but face smaller labeled datasets and require rigorous uncertainty quantification. Learning from AI's cycles of optimism and disappointment, success demands balancing genuine advances against hype.
\end{enumerate}
\end{summary}

\begin{issues}[FUTURE ISSUES]
\begin{enumerate}
\item Development of data-efficient deep learning methods for astronomy's label-scarce environment, leveraging physical priors and self-supervised strategies. Incorporation of unexplored physical symmetries into equivariant architectures.

\item Exploration of balancing physical constraints---conservation laws, differential equations---with discovery potential, determining which principles to encode as hard architectural constraints versus soft regularization for better generalization while maintaining physical consistency.

\item Broader implementation of multiscale neural modeling to bridge resolution gaps in simulations, learning effective subgrid prescriptions from high-fidelity runs for large-volume simulations. Discovering sparse latent representations could reveal interpretable features and emergent phenomena hidden in complex simulations.

\item Rigorous blind testing to determine whether field-level inference extracts more information than summary statistics in cosmology. SBI remains underutilized in complex astrophysical systems---exoplanets, stellar atmospheres, galaxy dynamics---where non-Gaussian features should provide clearer gains.

\item Development of rigorous frameworks and guarantees for neural density estimators in high-dimensional spaces to establish when SBI can be trusted. Proper uncertainty quantification through deep learning pipelines remains challenging yet essential. Standardized benchmarks with blind tests are needed, moving beyond theoretical exercises to real observations.

\item Exploration of emerging neural density estimation methods like flow matching that combine architectural flexibility with efficiency. More physically principled formulations with fewer barriers could accelerate adoption.

\item Development of foundation models demonstrating genuine zero-shot and few-shot transfer across wavelengths and instruments, with rigorous benchmarking against supervised methods to quantify information trade-offs. Focus needed on applications requiring domain transfer.

\item Expanded adoption of RL for operational astronomy including adaptive scheduling, instrument control, and observation optimization. Although demonstrated in controlled settings, deployment at observatories remains limited. Applications for algorithm optimization and simulation steering remain underexplored.

\item Ecosystem development enabling LLM agents to interact meaningfully with astronomical data through proper tools. Reorganizing existing toolkits into queryable, composable units could enable more autonomous workflows.

\item Education promoting deep learning as an extension of statistical methods rather than a replacement. Balancing progress against hype requires nuanced understanding of both capabilities and limitations across the community.
\end{enumerate}
\end{issues}

\section*{DISCLOSURE STATEMENT}
The author is not aware of any affiliations, memberships, funding, or financial holdings that might be perceived as affecting the objectivity of this review.

The author acknowledges the use of Claude (versions 3.7 Opus and 4.1 Opus) for copy-editing, improving text readability, and verifying accurate representation of cited work. Deep Research capability was used to identify potentially overlooked references. All conceptual development, structural organization, scientific interpretation, and reference selection remain entirely the author's work.

\section*{ACKNOWLEDGMENTS}
The author thanks Christopher Kochanek, David Weinberg, Hans-Walter Rix, and Xiaohui Fan for feedback on this review and acknowledges valuable discussions with colleagues and students who have shaped this understanding of the field.

\clearpage

\section*{SUPPLEMENTARY MATERIAL}

\section*{A. The Limitation of Classical Machine Learning Methods}
\label{app:classical-ml-comparison}

Deep learning has gained traction in astronomy because classical approaches often struggle to simultaneously achieve the three properties essential for modern astronomical challenges: scalability to large datasets, data efficiency when observations are limited, and expressivity to capture complex physical relationships. Understanding how classical methods navigate these trade-offs provides context for why physics-informed deep learning architectures have become increasingly attractive.

Linear regression and its variants, for instance, scale well through analytic solutions and learn effectively from minimal data \citep{Bishop2006,Ting2025a}, but their linear assumptions limit expressivity when modeling complex nonlinear astronomical phenomena.

Gaussian processes (GPs) represent the opposite trade-off. They offer expressivity through flexible kernel functions and provide principled Bayesian uncertainty quantification \citep{Rasmussen2006}. However, their computational complexity makes them intractable for large datasets, which is increasingly common in modern astronomical surveys.

\begin{figure}[h]
\centering
\includegraphics[width=\textwidth]{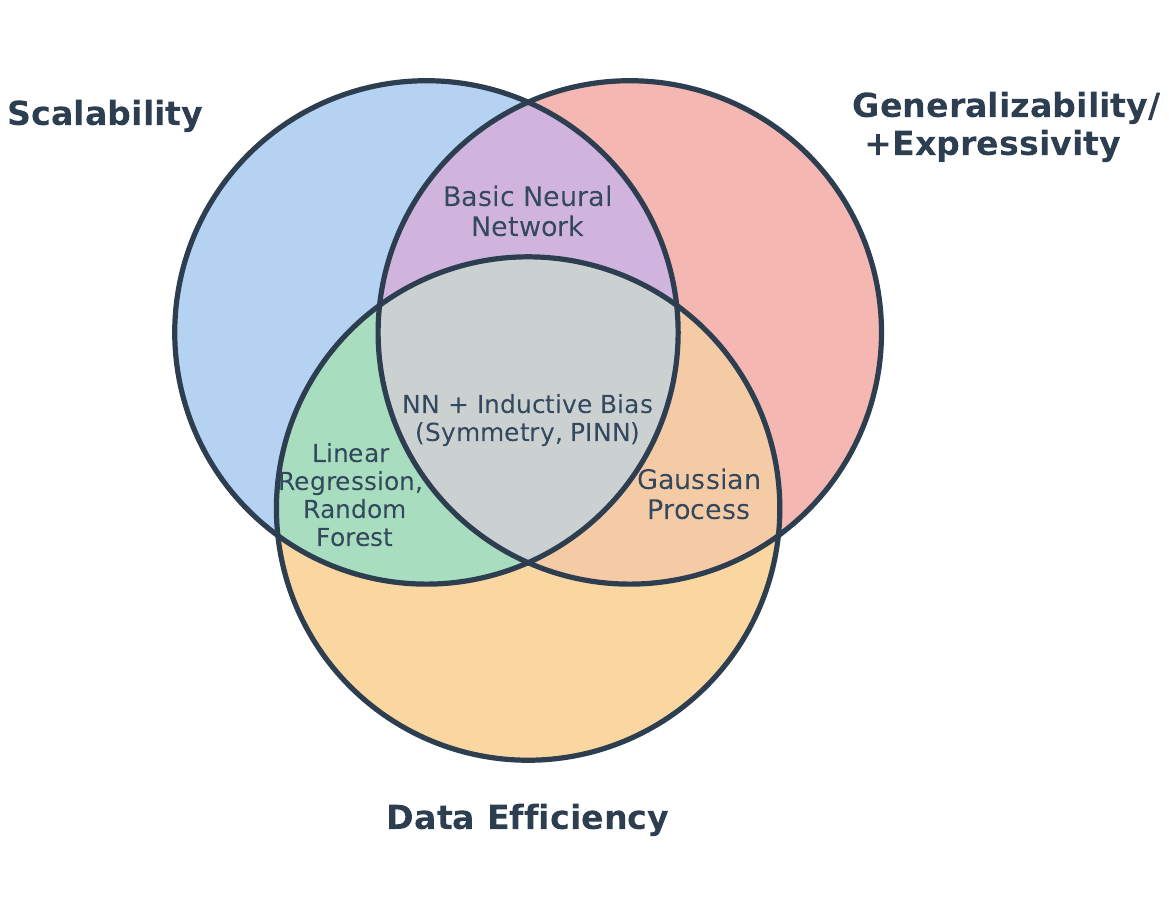}
\caption{Comparison of classical machine learning methods across three key properties for astronomical modeling. Gaussian processes excel in data efficiency and expressivity but lack scalability. Linear regression and random forests scale well and are data efficient but have limited expressivity and cannot extrapolate beyond training data. Basic neural networks offer scalability and expressivity but require large training datasets. Physics-informed deep learning aims to achieve all three properties by augmenting neural networks with appropriate physical constraints and symmetries.}
\label{fig4}
\end{figure}

Random forests and ensemble methods occupy a middle ground, balancing computational efficiency with nonlinear expressivity \citep{Breiman2001}. Their tree-based partitioning allows them to handle complex patterns while maintaining reasonable computational costs. However, their fundamental architecture restricts predictions to weighted combinations of training samples, preventing true extrapolation beyond the training data distribution---a critical limitation when modeling physical systems across unexplored parameter regimes.

\textbf{Figure~\ref{fig4}} visualizes these trade-offs across the three-dimensional property space, illustrating why no single classical method satisfies all requirements simultaneously and motivating the development of physics-informed deep learning architectures discussed in the main text.

\section*{B. Evolution of Deep Learning in Astrophysics}
\label{app:evolution-deep-learning}

Neural networks are not new to astronomy---applications date to the late 1980s \citep[e.g.,][]{Adorf1988,Angel1990,Odewahn1992,Storrie-Lombardi1992}. However, their nature and sophistication has changed considerably over the past decade. What began as simple function approximators have evolved into specialized architectures that encode physical assumptions and domain knowledge. This evolution from generic tools to physics-informed methods provides context for understanding why deep learning has become central to modern astronomical research.

The taxonomy presented here leverages large language models to analyze the astrophysics literature on astro-ph, using the same methodology that informed the broad conceptual landscape in \textbf{Figure~\ref{fig1}} of the main text. Building on \citet{Sun2024}, paper summaries from arXiv astro-ph (up to the end of July 2025) were analyzed to extract methodological concepts and construct a knowledge graph that captures technique relationships more robustly than keyword searches. Although \textbf{Figure~\ref{fig1}} illustrated high-level methodological categories across astronomy, the analysis here examines fine-grained adoption patterns of specific deep learning architectures---tracking the rise of individual techniques like convolutional neural networks, Transformers, and normalizing flows---to reveal how the field has selectively integrated specialized methods over time.

The complete knowledge graphs, citation networks, and extracted concepts are publicly available at \url{https://github.com/tingyuansen/astro-ph_knowledge_graph}.

\subsection*{B.1 Early neural networks in astronomy (1988--2015)}

In 2015, neural networks in astronomy consisted primarily of the most basic forms---multilayer perceptrons (feedforward neural networks)---and their simple variant, self-organizing maps. The former served as function approximators, the latter as dimension-reduction and clustering tools. Self-organizing maps \citep[SOMs;][]{Kohonen1982} create topology-preserving 2D grids in which similar inputs map to nearby nodes, effective for visualizing complex data distributions. SOMs found utility in classification tasks, from star/galaxy separation \citep{Mahonen1995} to light curve classification \citep{Brett2004}.

\begin{figure}[ht!]
    \centering
    \begin{subfigure}[b]{1.0\textwidth}
        \centering
        \caption*{{\bf Machine Learning Taxonomy in Astronomy in 2015}}
        \includegraphics[width=\textwidth]{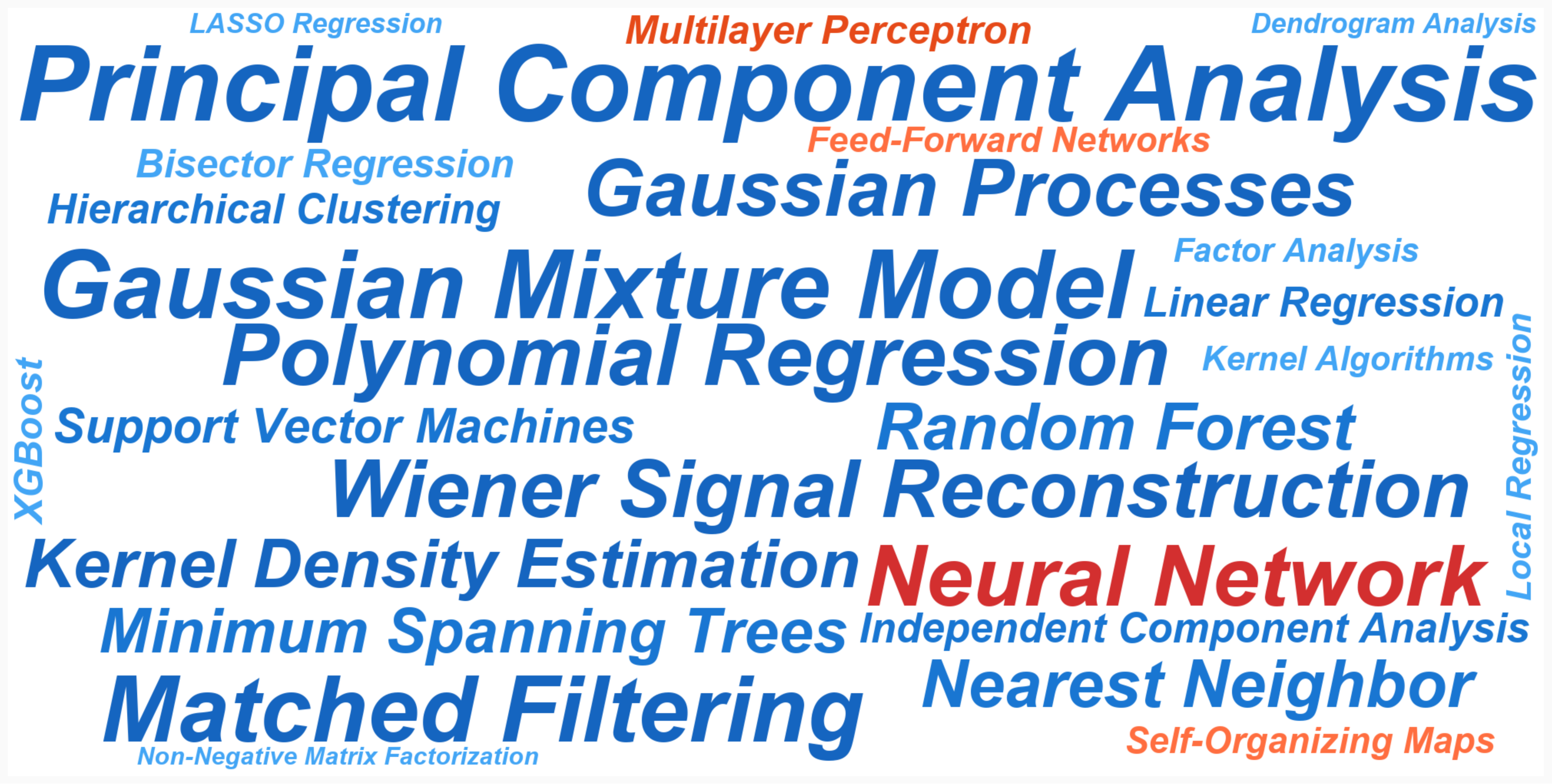}
    \end{subfigure}
    \vspace{0.5cm}
    \begin{subfigure}[b]{1.0\textwidth}
        \centering
        \caption*{{\bf Machine Learning Taxonomy in Astronomy in 2025}}
        \includegraphics[width=\textwidth]{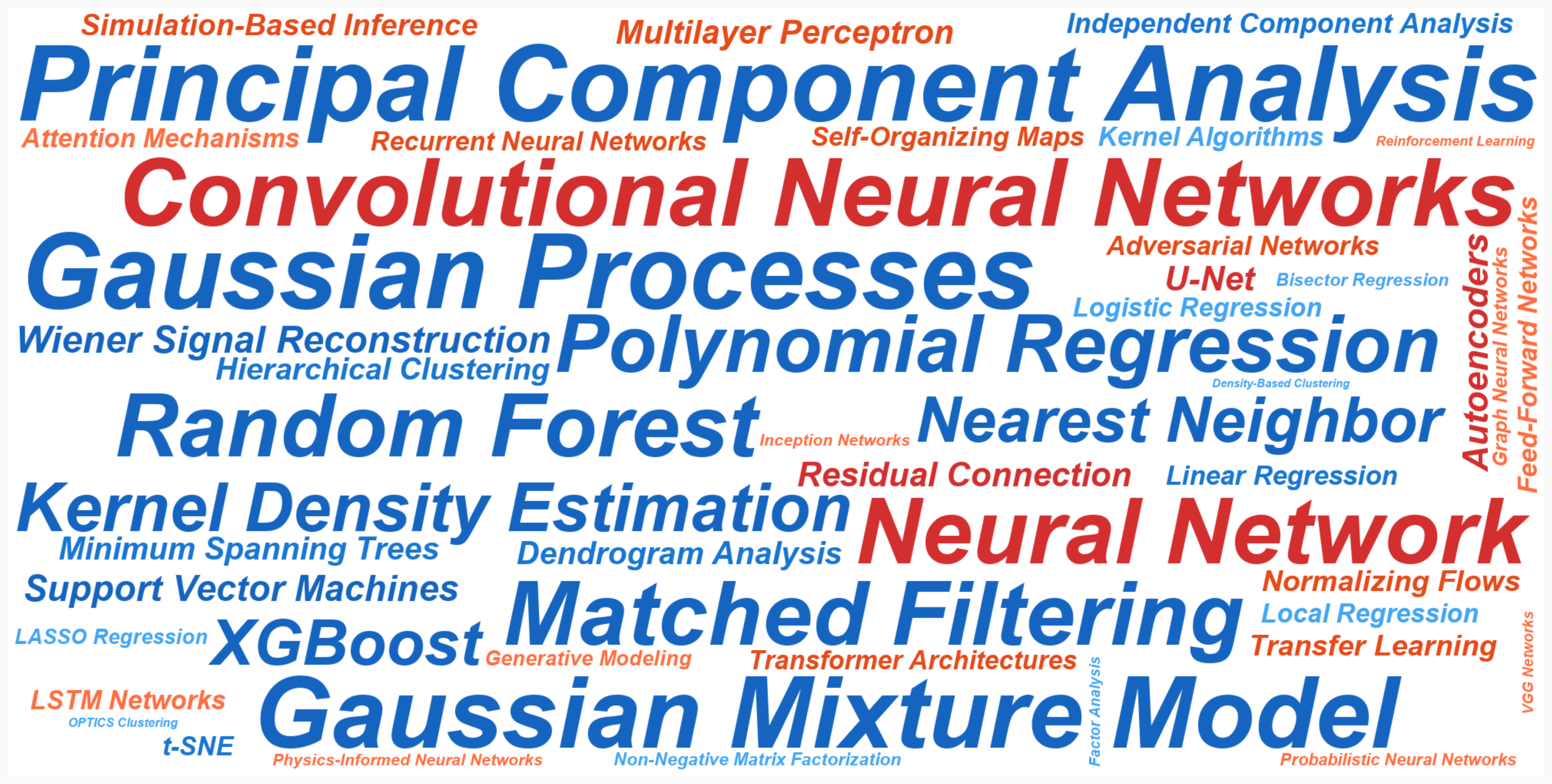}
    \end{subfigure}
    \caption{Evolution of machine learning methods in astronomy. (\textit{Top}) The 2015 taxonomy shows neural networks primarily as flexible interpolators (multilayer perceptrons, feedforward networks) alongside classical methods. (\textit{Bottom}) The 2025 taxonomy shows neural network approaches constitute a substantial fraction of applications, with specialized architectures (convolutional neural networks, Transformers, graph neural networks) and probabilistic methods (normalizing flows, simulation-based inference, variational autoencoders) reflecting maturation beyond simple function approximation.}
    \label{fig5}
\end{figure}

\textbf{Figure~\ref{fig5}} visualizes this contrast between 2015 and 2025. The 2015 taxonomy (top panel) shows neural networks as one component among many classical methods, occupying relatively modest space in the methodological landscape. Neural networks and self-organizing maps appear alongside dominant classical techniques like Gaussian processes, principal component analysis, and random forest. Neural networks primarily served as flexible interpolators rather than specialized tools.

By 2025 (bottom panel), the landscape has changed considerably, reflecting widespread adoption. The methodological space has diversified: specialized architectures like convolutional neural networks, Transformer architectures, U-Net, autoencoders, and graph neural networks appear more prominently. This shift reflects the field's move toward architectures that encode specific assumptions about data structure and physical constraints. Probabilistic deep learning methods---normalizing flows, simulation-based inference, variational autoencoders---indicate maturation beyond simple function approximation.

\subsection*{B.2 The rise of specialized architectures (2016--2025)}

\begin{figure}[ht!]
    \centering
    \includegraphics[width=1.0\textwidth]{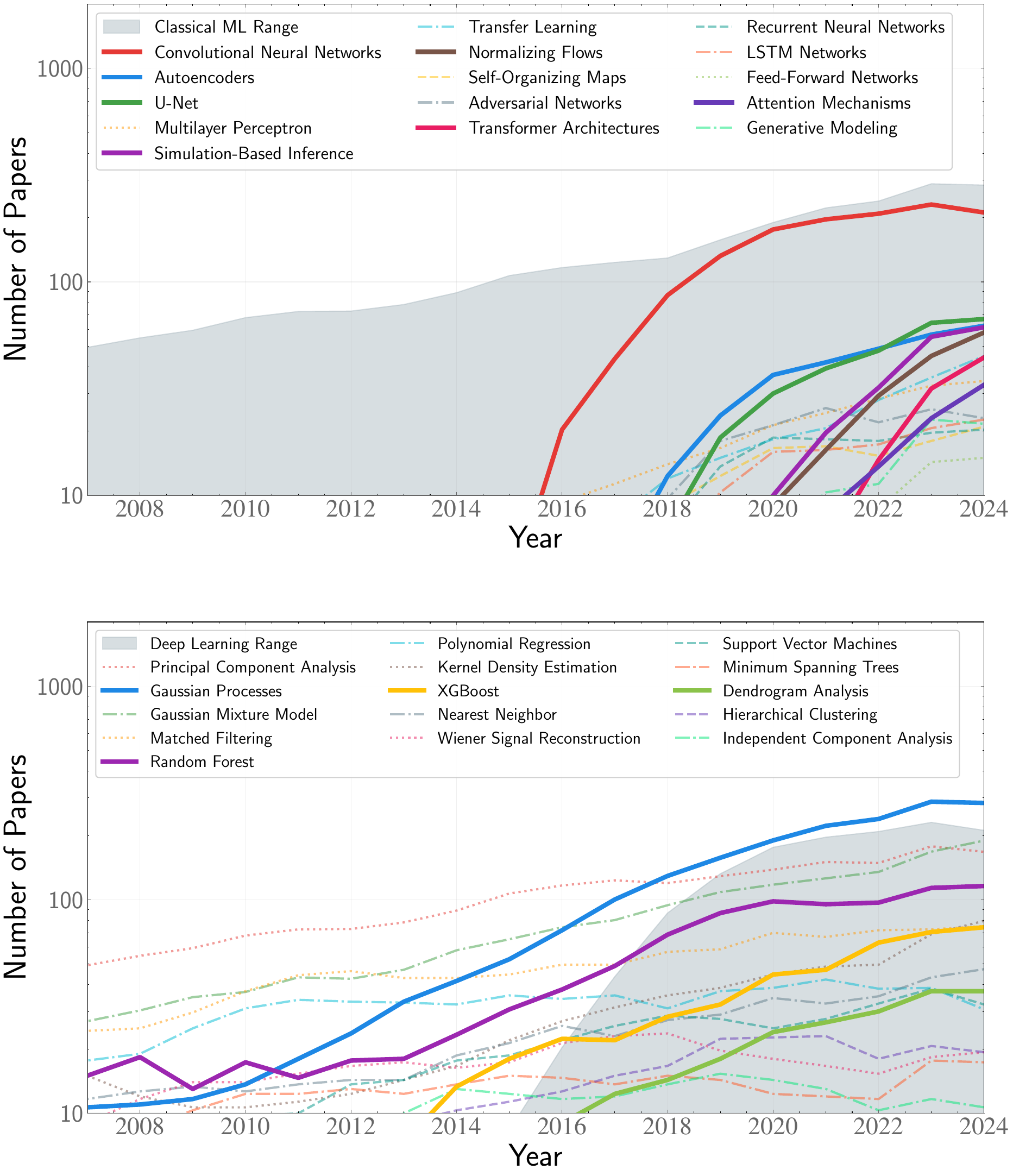}
    \caption{Growth trajectories of machine learning techniques in astronomy. (\textit{Top}) Deep learning methods. Convolutional neural networks show adoption from 2016, followed by U-Net for segmentation (2018), autoencoders/VAEs for dimensionality reduction (2018), simulation-based inference with normalizing flows for Bayesian modeling (2020), and Transformers and attention mechanism showing rapid growth after 2022. (\textit{Bottom}) Classical techniques showing sustained adoption. Grey shading indicates the envelope of techniques from the other category/panel for reference.}
    \label{fig6}
\end{figure}

\textbf{Figure~\ref{fig6}} reveals the adoption trajectories of individual techniques from 2008 to 2025, exposing the temporal dynamics behind the taxonomic shift. The top panel tracks deep learning methods, whereas the bottom panel shows classical techniques for comparison, with grey shading indicating the envelope from the opposite category.

Convolutional neural networks led the deep learning wave, showing exponential growth starting around 2016. This timing follows breakthrough computer vision architectures: InceptionNet \citep{Szegedy2014}, VGGNet \citep{Simonyan2014}, and ResNet \citep{He2016}. Around 2018, a second wave emerged. U-Net \citep{Ronneberger2015}, designed for pixel-level predictions, found application in artifact removal \citep[e.g.,][]{Zhang2020} and source segmentation \citep[e.g.,][]{Hausen2020}. Autoencoders and variational autoencoders \citep[VAEs;][]{Kingma2013} rose concurrently as alternatives to traditional dimensionality reduction like principal component analysis \citep{Baldi1989}, offering nonlinear latent representations that better capture complex data manifolds.

The most recent wave began around 2020--2022. Simulation-based inference (SBI), powered by normalizing flows \citep{JiminezRezende2015}, enables rigorous Bayesian inference matching astronomy's statistical traditions while scaling to modern data volumes (discussed in Section \ref{subsection:simulation-based-inference}). Transformer architectures \citep{Vaswani2017} show rapid growth after 2022, excelling at capturing long-range dependencies in irregular time series and spectroscopic data through attention mechanisms (Section \ref{subsubsection:transformer-architectures}).

\subsection*{B.3 Field-wide adoption patterns}

\begin{figure}[ht!]
    \centering
    \includegraphics[width=\textwidth]{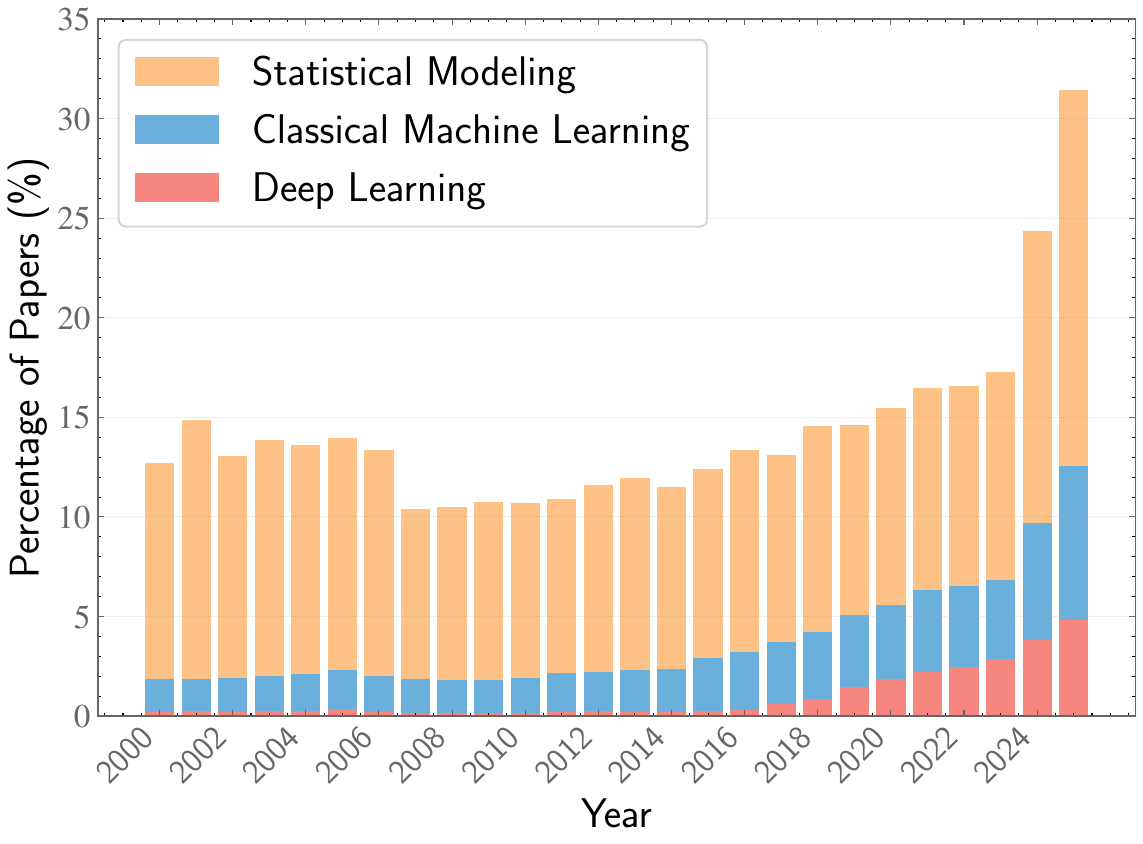}
    \caption{Temporal evolution of statistical methods in astronomy (2000--July 2025) from astro-ph literature. Shows fraction of papers employing concepts in statistical modeling, classical machine learning, and deep learning. Papers categorized by highest-level methodology (deep learning $>$ classical machine learning $>$ statistical modeling). Deep learning rises from negligible presence to 5\% of papers by 2025, catching up with classical machine learning adoption. The total fraction of papers using advanced quantitative methods (combining all three categories) has increased from 10--15\% in 2000 to 30\% by 2025, indicating that statistical approaches generally are becoming more central to astronomical research.}
    \label{fig7}
\end{figure}

These individual technique trajectories aggregate into a broader pattern of methodological evolution. \textbf{Figure~\ref{fig7}} shows the fraction of astro-ph papers employing statistical modeling, classical machine learning, and deep learning from 2000 to July 2025. Papers are categorized by their highest-level methodology (deep learning $>$ classical machine learning $>$ statistical modeling), meaning a paper using both deep learning methods (e.g., CNNs) and classical machine learning techniques (e.g., random forests) in the same paper would be classified as deep learning.

From 2000 to 2015, the landscape remained relatively stable. Statistical modeling held at 10--15\% of papers, whereas classical machine learning remained at 2\%. Deep learning was essentially absent. The inflection point occurs around 2016--2018, coinciding with the CNN adoption wave shown in \textbf{Figure~\ref{fig6}}. Deep learning adoption accelerates exponentially, reaching 5\% of papers by 2025, matching classical machine learning adoption levels.

The total fraction employing advanced quantitative methods---combining statistical modeling, classical machine learning, and deep learning---has grown from 10--15\% in 2000 to 30\% by 2025. This increase indicates that statistical approaches generally are becoming more central to astronomical research.

\section*{C. Symmetry Encoding and Physics Constraints}
\label{app:symmetry-encoding-illustration}

\begin{figure}[ht!]
\centering
\includegraphics[width=1.0\textwidth]{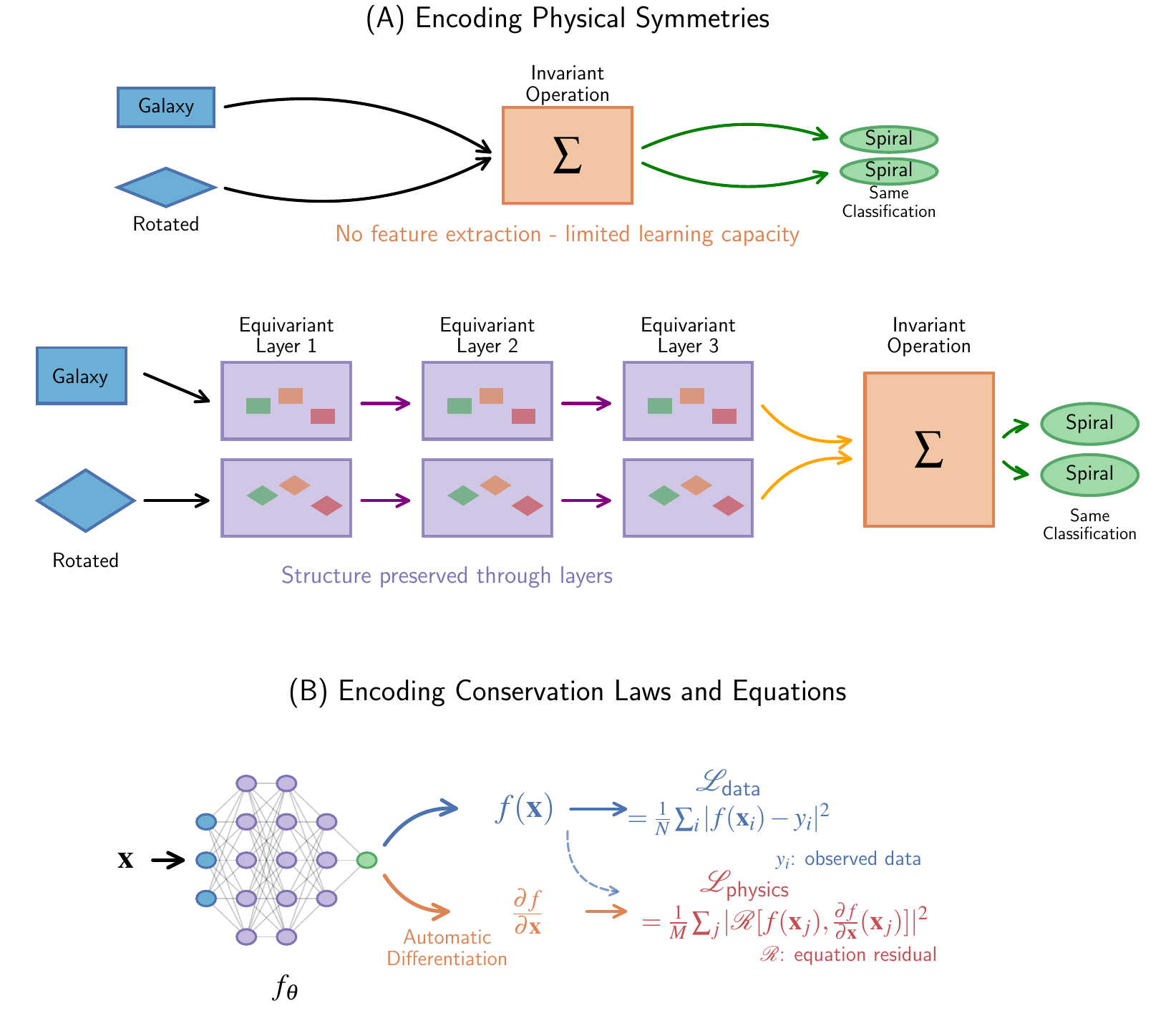}
\caption{Two complementary approaches for encoding physical knowledge into neural networks. (\textit{A}) Encoding physical symmetries: Equivariant layers preserve geometric structure through feature extraction---rotated galaxies produce correspondingly rotated feature maps---before a final invariant operation produces consistent classifications. This maintain-structure-then-summarize principle enables networks to learn expressive features while respecting physical symmetries. (\textit{B}) Encoding conservation laws and equations: Physics-informed neural networks simultaneously fit observational data and satisfy governing equations through automatic differentiation. The network learns solutions that respect both empirical observations ($\mathcal{L}_{\text{data}}$) and physical constraints ($\mathcal{L}_{\text{physics}}$), enabling extrapolation beyond training data through encoded domain knowledge.}
\label{fig8}
\end{figure}

Neural networks can encode physical knowledge through two complementary approaches: symmetry preservation through equivariant architectures, and explicit physical constraints through equation-informed training. \textbf{Figure~\ref{fig8}} illustrates both strategies.

\subsection*{C.1 Encoding physical symmetries through equivariance}

The principle of equivariant feature extraction followed by invariant prediction can be understood by contrasting two approaches. A naive approach would apply invariant operations immediately---for example, computing rotation-invariant summary statistics directly from input images. Although this produces consistent outputs for rotated inputs, it discards geometric structure at the outset, severely limiting the network's ability to learn expressive features.

The equivariant approach instead maintains geometric structure through intermediate layers. When processing a galaxy image, equivariant convolutional layers produce feature maps that rotate correspondingly with the input. A 90-degree rotation of the input galaxy produces a 90-degree rotation of all intermediate feature representations. Only after extracting rich, structure-preserving features does a final global operation (such as spatial averaging or max pooling) produce the invariant output needed for classification or regression.

\textbf{Figure~\ref{fig8}} (top and middle panels) illustrates this distinction. The equivariant pathway preserves spatial relationships throughout feature extraction, enabling the network to learn complex patterns while respecting physical symmetries. This maintain-structure-then-summarize principle proves far more expressive than immediate invariance, as it allows the network to leverage geometric information during learning before producing the required invariant predictions.

\subsection*{C.2 Encoding conservation laws and equations}

Beyond symmetries, neural networks can encode explicit physical constraints through their loss functions. Physics-informed neural networks simultaneously satisfy observational data and governing equations by combining two loss terms: a data fidelity term $\mathcal{L}_{\text{data}}$ that measures fit to observations, and a physics term $\mathcal{L}_{\text{physics}}$ that penalizes violations of known equations.

The physics loss is computed using automatic differentiation to evaluate whether the network's predictions satisfy differential equations. For example, if modeling stellar structure, the network must produce temperature and pressure profiles that satisfy hydrostatic equilibrium. The total loss $\mathcal{L} = \mathcal{L}_{\text{data}} + \lambda \mathcal{L}_{\text{physics}}$ ensures the learned solution respects both empirical observations and physical laws.

\textbf{Figure~\ref{fig8}} (bottom panel) illustrates this dual constraint approach. By encoding domain knowledge as differentiable constraints, the network can extrapolate beyond training data more reliably than purely data-driven models. This framework proves particularly valuable when data are sparse but governing equations are known, allowing physical laws to regularize the learned solution.

\section*{D. Neural Density Estimation for Simulation-Based Inference and Anomaly Detection}
\label{app:neural-density-estimation}

Neural density estimation methods enable both simulation-based inference (SBI) and anomaly detection in astronomy. This section illustrates the architectural differences between normalizing flows and diffusion models, explains the distinction between neural likelihood estimation (NLE) and neural posterior estimation (NPE), and demonstrates how the same density estimation framework supports both parameter inference and outlier detection.

\subsection*{D.1 Normalizing flows and diffusion models: architectural comparison}

\begin{figure}[ht!]
\centering
\includegraphics[width=1.0\textwidth]{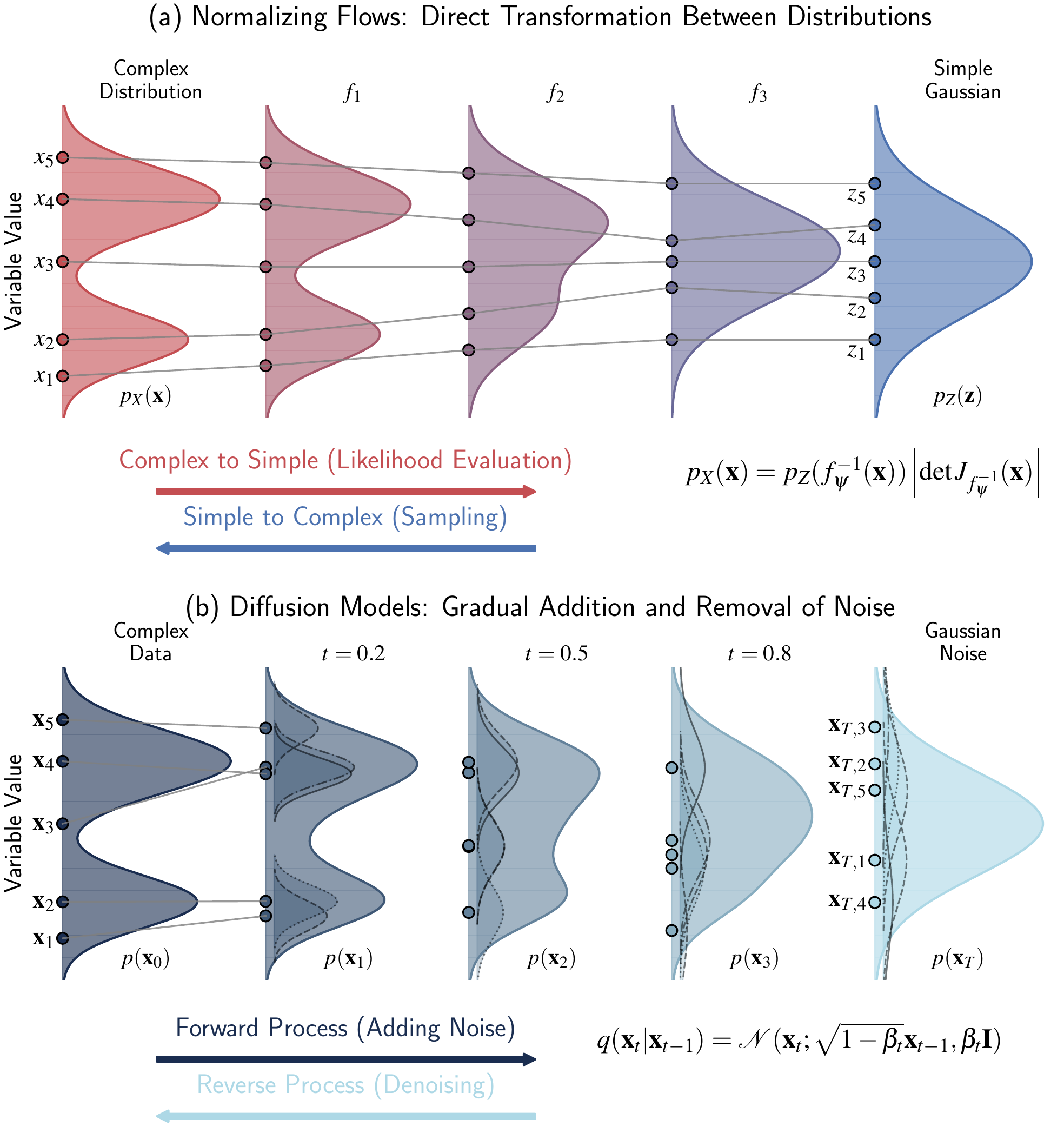}
\caption{Neural density estimation approaches for simulation-based inference. (\textit{a}) Normalizing flows: Direct invertible transformations between complex distributions $p_X(\mathbf{x})$ and simple Gaussians $p_Z(\mathbf{z})$ via learned functions $f_\psi$, enabling exact likelihood computation through the change of variables formula. The bidirectional arrows indicate that the transformation works both ways: evaluating densities (complex to simple) and generating samples (simple to complex). (\textit{b}) Diffusion models: Gradual noise corruption process over time steps $t$, transforming complex data distributions to Gaussian noise $\mathcal{N}(0, \mathbf{I})$. The neural network learns to reverse this process, enabling sample generation by iterative denoising. Figure adapted with permission from \citet{Ting2025a}.}
\label{fig9}
\end{figure}

Normalizing flows perform direct transformations between complex distributions and simple Gaussians through invertible functions. The upper panel of \textbf{Figure~\ref{fig9}} illustrates this approach: The network learns a bijective mapping $f_\psi$ that transforms a complex data distribution $p_X(\mathbf{x})$ into a standard Gaussian $p_Z(\mathbf{z})$. The change of variables formula provides exact likelihood computation,
\begin{equation}
p_X(\mathbf{x}) = p_Z(f_\psi^{-1}(\mathbf{x}))|\det J_{f_\psi^{-1}}(\mathbf{x})|.
\end{equation}
This direct transformation enables fast inference through a single forward pass. However, invertibility with tractable Jacobians constrains network architecture.

Diffusion models take a gradual approach. The lower panel of \textbf{Figure~\ref{fig9}} shows the forward process that progressively adds Gaussian noise over many time steps $t$:
\begin{equation}
q(\mathbf{x}_t|\mathbf{x}_{t-1}) = \mathcal{N}(\mathbf{x}_t; \sqrt{1-\beta_t}\mathbf{x}_{t-1}, \beta_t\mathbf{I}).
\end{equation}
After sufficient steps, the distribution converges to $\mathcal{N}(0, \mathbf{I})$. The neural network learns to reverse this corruption by iterative denoising. This relaxes architectural constraints and improves training stability, but requires hundreds of network evaluations per sample, making diffusion models much slower than normalizing flows for inference.

\begin{figure}[ht!]
\centering
\includegraphics[width=\textwidth]{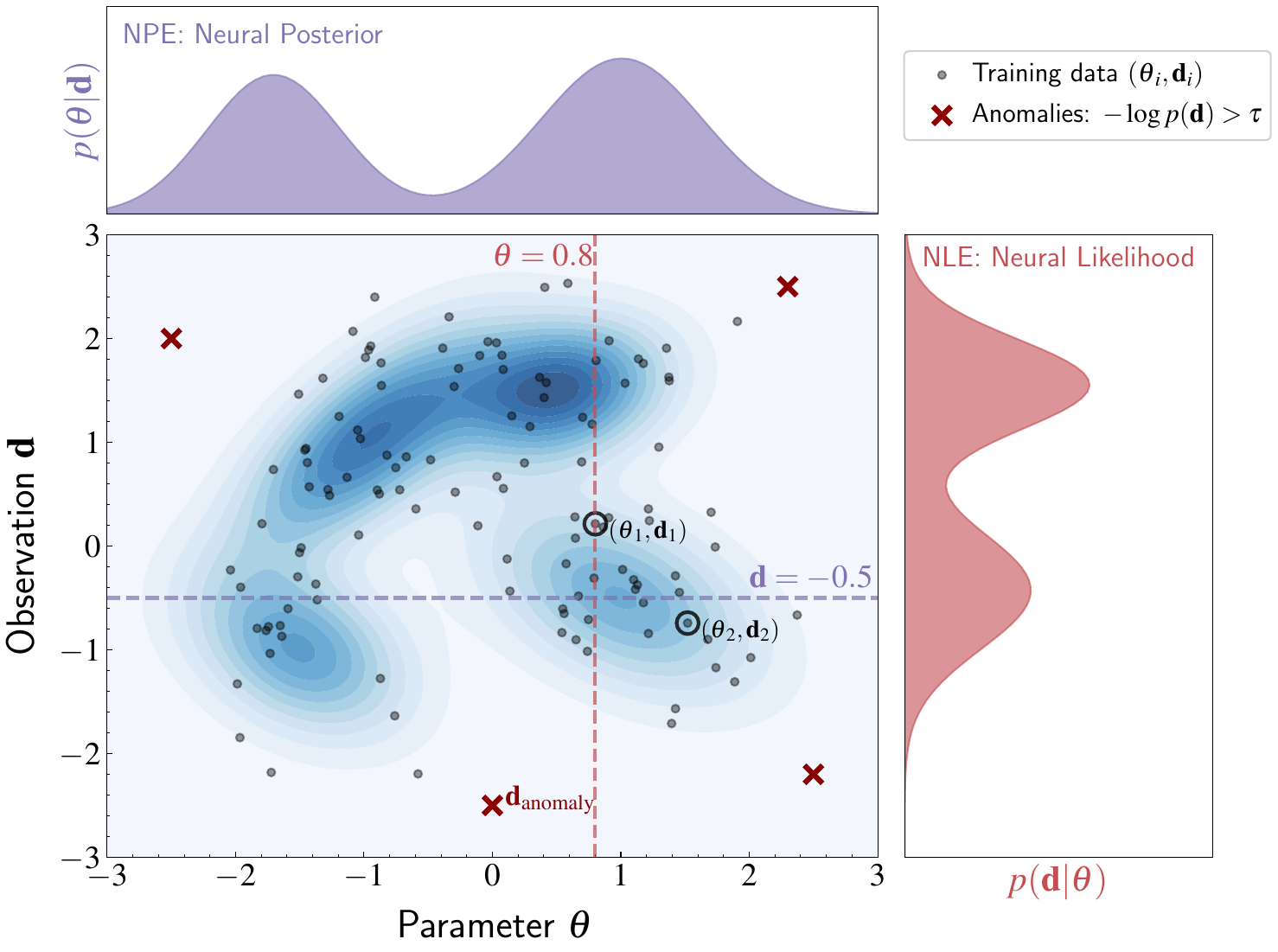}
\caption{Neural density estimation for simulation-based inference and anomaly detection. Neural networks can learn either the joint distribution $p(\boldsymbol{\theta}, \mathbf{d})$, the posterior $p(\boldsymbol{\theta}|\mathbf{d})$ directly (NPE), or the likelihood $p(\mathbf{d}|\boldsymbol{\theta})$ (NLE) from forward simulations, where $\boldsymbol{\theta}$ represents model parameters and $\mathbf{d}$ represents observational data. Both $\boldsymbol{\theta}$ and $\mathbf{d}$ can be very high-dimensional; $\mathbf{d}$ can be summary statistics or, in the case of field-level inference, entire images, spectra, or other modalities. Black points show training data pairs $(\boldsymbol{\theta}_i, \mathbf{d}_i)$ generated from simulations or obtained from empirical data. The figure also illustrates the connection to anomaly detection: Red crosses mark anomalies in which likelihood is lower than a chosen threshold $\tau$, demonstrating how the same density estimation framework enables both Bayesian parameter inference and systematic outlier detection in astronomical surveys. Figure adapted with permission from \citet{Ting2025a}.}
\label{fig10}
\end{figure}

\subsection*{D.2 Neural likelihood estimation versus neural posterior estimation}

Two strategies exist for Bayesian inference using neural density estimators: neural likelihood estimation (NLE) and neural posterior estimation (NPE). \textbf{Figure~\ref{fig10}} illustrates both approaches. Both methods train neural networks on simulation-observation pairs $(\boldsymbol{\theta}_i, \mathbf{d}_i)$ generated by sampling parameters from the prior and running forward simulations.

Recall that Bayesian inference seeks the posterior distribution:
\begin{equation}
p(\boldsymbol{\theta}|\mathbf{d}) = \frac{p(\mathbf{d}|\boldsymbol{\theta}) p(\boldsymbol{\theta})}{p(\mathbf{d})},
\end{equation}
where $p(\mathbf{d}|\boldsymbol{\theta})$ is the likelihood, $p(\boldsymbol{\theta})$ is the prior, and $p(\mathbf{d})$ is the evidence.

NLE learns the likelihood $p(\mathbf{d}|\boldsymbol{\theta})$ and requires Markov chain Monte Carlo (MCMC) sampling to obtain posteriors by applying Bayes' theorem. This offers flexibility: One can change priors without retraining, perform model comparison, and combine multiple observations. However, MCMC adds computational cost for each inference.

NPE directly learns the posterior $p(\boldsymbol{\theta}|\mathbf{d})$, providing immediate parameter constraints without additional sampling. This amortization makes NPE efficient for repeated inference on many observations. The trade-off is reduced flexibility: Changing the prior requires retraining, and model comparison becomes more challenging.

The choice depends on scientific requirements. NLE suits scenarios with uncertain priors or requiring model comparison. NPE excels when priors are well-established and inference speed is critical, such as real-time survey operations.

\subsection*{D.3 Anomaly detection through density estimation}

The same neural density estimators used for SBI naturally enable systematic anomaly detection. By learning the distribution $p(\mathbf{d})$ of normal astronomical objects, these methods quantify how typical or atypical any observation appears. \textbf{Figure~\ref{fig10}} illustrates this connection. The red crosses mark anomalies in which the likelihood falls below a chosen threshold $\tau$: objects for which $-\log p(\mathbf{d}) > \tau$. This threshold can be set to control the false positive rate based on the desired sensitivity. This approach provides a principled probabilistic measure rather than arbitrary distance metrics. Furthermore, it requires only normal training data without labeled anomalies, making it valuable for discovering unknown classes of astronomical phenomena.

\begin{figure}[ht!]
\centering
\includegraphics[width=1.0\textwidth]{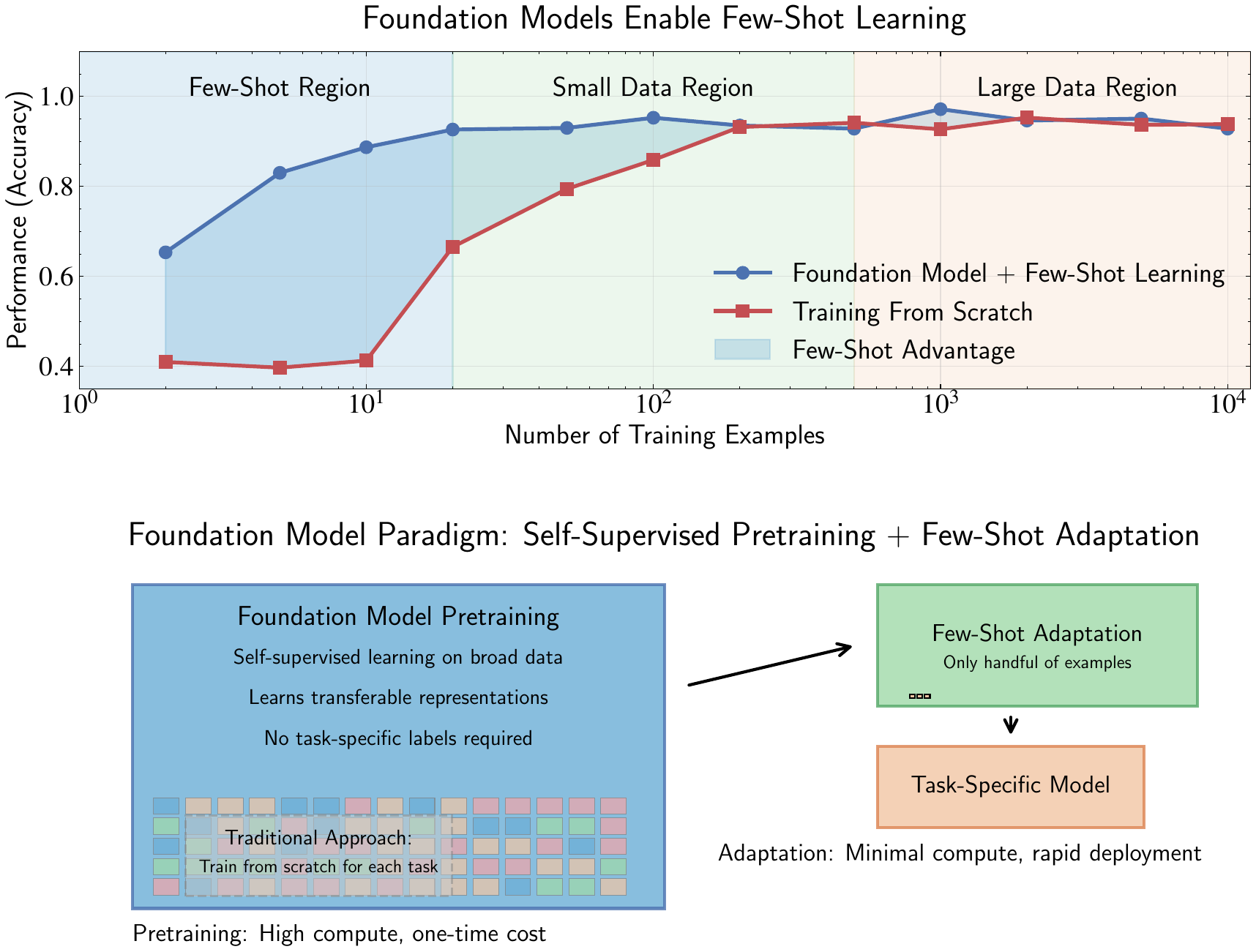}
\caption{Foundation models versus traditional supervised learning. Traditional supervised learning (red) requires substantial labeled datasets to achieve good performance, with each new task starting from scratch. Foundation models (blue) achieve strong performance with minimal task-specific data by pretraining on diverse unlabeled data through self-supervised objectives. The shaded region indicates where foundation models provide greatest advantage: tasks with scarce labeled data but abundant unlabeled data. Figure adapted with permission from \citet{Ting2025a}.}
\label{fig11}
\end{figure}

\section*{E. Foundation Models: Learning Curves and Data Efficiency}
\label{app:foundation-models}

Foundation models aim to achieve strong performance with minimal task-specific labeled data by leveraging self-supervised pretraining on large unlabeled datasets. \textbf{Figure~\ref{fig11}} contrasts this paradigm with traditional supervised learning. Traditional methods (red curve) require substantial labeled datasets to achieve good performance, with each new task starting from scratch---the learning curve begins at low performance and gradually improves only with extensive task-specific data. Foundation models (blue curve) follow a fundamentally different trajectory: Pretraining on diverse unlabeled data enables strong initial performance that rapidly improves with even minimal labeled examples. The shaded region highlights where foundation models provide greatest advantage: tasks with scarce labeled data but abundant unlabeled data, a common scenario in astronomy.

For example, foundation models could address the synthetic-to-observation gap in stellar spectroscopy. Full 3D magnetohydrodynamic stellar atmosphere simulations with non-LTE radiative transfer remain computationally prohibitive at scale. Consequently, practical analysis relies on 1D models assuming LTE \citep{Kurucz1993}---millions of times faster but missing crucial physics like 3D convection and nonequilibrium effects. This gap becomes critical for unique observations with limited calibration data, such as interpreting M31 stellar spectra from JWST \citep{Nidever2024}.

Foundation models offer a potential solution, though largely aspirational at present. The approach would pretrain on massive grids of simplified 1D LTE simulations, fine-tune on limited 3D non-LTE simulations \citep{Freytag2012}, then apply to real observations. The underlying physics remains consistent---how temperature stratification creates absorption profiles, how pressure broadening depends on density, how abundances manifest in line strengths. A foundation model could learn these relationships from simple models, refine them with complex simulations, and apply them to observations, potentially unlocking analyses currently impossible with traditional approaches.

\clearpage

\nocite{Adorf1988}
\nocite{Angel1990}
\nocite{Baldi1989}
\nocite{Bishop2006}
\nocite{Breiman2001}
\nocite{Brett2004}
\nocite{Freytag2012}
\nocite{Hausen2020}
\nocite{He2016}
\nocite{JiminezRezende2015}
\nocite{Kingma2013}
\nocite{Kohonen1982}
\nocite{Kurucz1993}
\nocite{Mahonen1995}
\nocite{Nidever2024}
\nocite{Odewahn1992}
\nocite{Rasmussen2006}
\nocite{Ronneberger2015}
\nocite{Simonyan2014}
\nocite{Storrie-Lombardi1992}
\nocite{Sun2024}
\nocite{Szegedy2014}
\nocite{Ting2025a}
\nocite{Vaswani2017}
\nocite{Zhang2020}

\bibliographystyle{ar-style2}  
\bibliography{manuscript}

\clearpage

\end{document}